\documentclass[11pt]{article}

\usepackage{amssymb,amsmath,latexsym}
\usepackage{graphicx}
\usepackage{jinstpub}
\usepackage{caption}
\usepackage{subcaption}
\usepackage{listings}
\usepackage{color}
\usepackage{xspace}
\usepackage{placeins}

\definecolor{dkgreen}{rgb}{0,0.6,0}
\definecolor{gray}{rgb}{0.5,0.5,0.5}
\definecolor{mauve}{rgb}{0.58,0,0.82}

\newcommand{\mb}{MiniBooNE\xspace}
\newcommand{\minerva}{MINERvA\xspace}

\newcommand{\mtlabel[1]}{\mbox{\tiny #1}}
\definecolor{mygreen}{rgb}{0,0.6,0}
\definecolor{mygray}{rgb}{0.5,0.5,0.5}
\definecolor{mymauve}{rgb}{0.58,0,0.82}

\notoc

\title{{\bf NUISANCE:} a neutrino cross-section generator tuning and comparison framework}

\collaboration{\includegraphics[height=30mm]{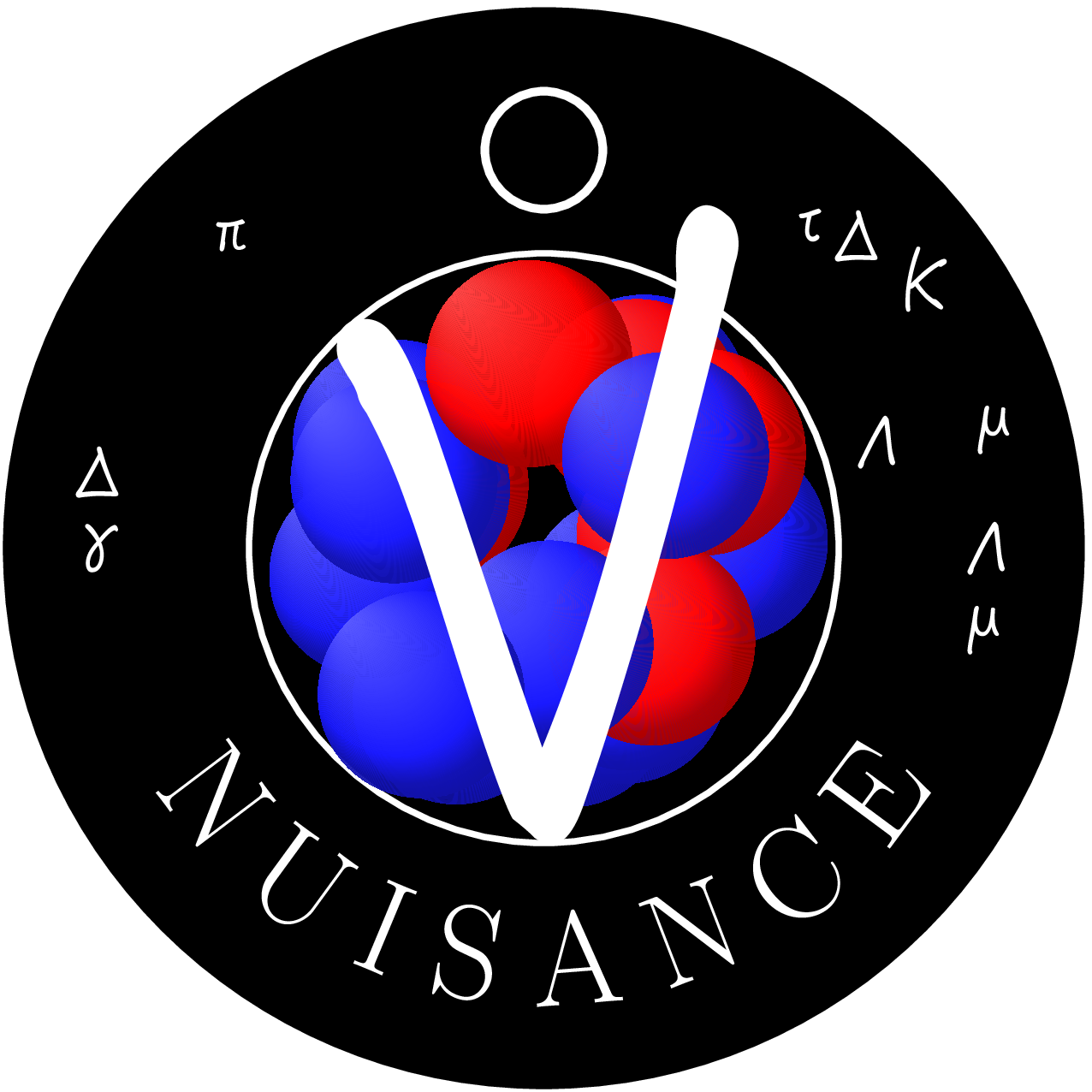}}

\date{\today}
\author[a]{P.~Stowell}
\author[b]{C.~Wret}
\author[c]{C.~Wilkinson}
\author[b]{L.~Pickering}
\author[a]{S.~Cartwright}
\author[d,e]{Y.~Hayato}
\author[f]{K.~Mahn}
\author[g]{K.S.~McFarland}
\author[h]{J.~Sobczyk}
\author[i]{R.~Terri}
\author[a]{L.~Thompson}
\author[b]{M.O.~Wascko}
\author[b]{Y.~Uchida}
\affiliation[a]{University of Sheffield, Department of Physics and Astronomy, Sheffield, United Kingdom}
\affiliation[b]{Imperial College London, Department of Physics, London, United Kingdom}
\affiliation[c]{University of Bern, Albert Einstein Center for Fundamental Physics, Laboratory for High Energy Physics (LHEP), Bern, Switzerland}
\affiliation[d]{Kavli Institute for the Physics and Mathematics of the Universe (WPI),Todai Institutes for Advanced Study, University of Tokyo, Kashiwa, Chiba, Japan}
\affiliation[e]{University of Tokyo, Institute for Cosmic Ray Research, Kamioka Observatory, Kamioka, Japan}
\affiliation[f]{Michigan State University, Department of Physics and Astronomy, East Lansing, Michigan, U.S.A}
\affiliation[g]{University of Rochester, Department of Physics and Astronomy, Rochester, New York, USA}
\affiliation[h]{University of Wroc\l{}aw, Institute of Theoretical Physics, Wroc\l{}aw, Poland}
\affiliation[i]{Queen Mary University of London, School of Physics and Astronomy, London, United Kingdom}
\emailAdd{nuisance@projects.hepforge.org}

\abstract{NUISANCE is an open source C++ framework which facilitates detailed studies of neutrino interaction cross-section models implemented in Monte Carlo neutrino event generators. It provides a host of automated methods to perform comparisons of multiple generators to published cross-section measurements and each other. External reweighting libraries are used to allow the end-user to evaluate the impact of model parameters variations in the generators with data, or to tune the generator predictions to arbitrary dataset combinations. The design is modular and focusses on ease-of-use to allow new datasets and more generators to be added without requiring detailed understanding of the entire NUISANCE package. We discuss the motivation for the NUISANCE framework and suggested usage cases, alongside a description of its core structure.} 
\begin{document}
\maketitle
\flushbottom

\section{Introduction}
Current and future neutrino oscillation experiments have stringent systematic error budgets, which are considerably smaller than are currently achievable. In particular, the uncertainties associated with neutrino interaction cross-section models need to be reduced down to the few percent level to deliver the required sensitivy; required cross-section uncertainties of $4\%$ and $2\%$ have been projected for T2K-II and DUNE respectively~\cite{lbne,t2k_sensitivity_2014}. Long baseline oscillation experiments spanning the $0.1 \leq E_{\nu} \leq 10$ GeV range suffer especially, as at these energies a consistent theoretical interaction picture has yet to emerge~\cite{zeller12, hayato_review_2014, garvey_review_2014}. Selecting default interaction models from those available and estimating parameter uncertainties are significant challenges currently facing neutrino oscillation and cross-section experiments.

Two main issues complicate the problem of building a consistent neutrino cross-section model when using \emph{nuclear targets}. Firstly, the interaction-level variables which cross-section models are constructed in terms of (e.g., energy and momentum transfer, neutrino energy) cannot be directly measured by experiments. The incoming neutrino four-momentum is not known on an event-by-event basis from the beam, nor can it be reconstructed accurately by using final-state particle kinematics without relying on the experiment's model for nuclear effects, such as initial state nucleon model and particle propagation. The only measurable \emph{model-independent} quantities are the outgoing particle kinematics (e.g., outgoing muon momentum and direction). Secondly, Final State Interactions (FSI)---where the particles leaving the interaction vertex re-interact before leaving the nucleus---can modify the outgoing particle kinematics and event particle content. It is not possible to separate a single interaction process with selection cuts: a simple $\nu_\mu + n \rightarrow \mu^{-} + p$ Charged-Current Quasi-Elastic (CCQE) interaction cannot be clearly distinguished from a $\nu_{\mu} + p \rightarrow \mu^{-} + \pi^{+} + p$ interaction if the pion is absorbed in the nucleus. As a result, model-independent measurements must, in general, describe a final-state \emph{topological} cross section rather than a single interaction mode cross section---such as measuring events with one muon and no pions in the final state (CC$0\pi$ interactions) in lieu of CCQE interactions.

A number of general purpose neutrino interaction Monte Carlo (MC) event generators are available, simulating a large range of interactions. These make it possible to produce realistic predictions for topological cross-section measurements, and allow the user to modify model parameters and combine different models. Whilst model-independent measurements are essential for arriving at a well-motivated cross-section model with defensible uncertainties, topological cross sections given in terms of final-state particle kinematics provide relatively weak constraints of cross-section model parameters which often have most strength in interaction variables such as four-momentum transfer. This is complicated by the fact that different detector sensitivities and kinematic thresholds mean experiments may probe vastly different regions of phase-space for an observed interaction channel. Therefore model parameters extrapolated from one experiment may not be sufficient to describe all other experiments.
Hence it is essential to use data from many experiments, with different energies, target materials and detector designs to constrain a full cross-section model and claim confidence in it.

NUISANCE is a software package written to simplify the task of confronting and comparing neutrino event generators with each other and published world cross-section data. It is an open source C++ package distributed under the GPLv3 license agreement~\cite{gplv3}. NUISANCE is the primary tool for evaluating and constraining the cross-section model used in T2K analyses~\cite{Wilkinson:2016wmz} using external scattering data, and grew out of efforts to tune the NEUT interaction model within the T2K Neutrino Interactions Working Group. The main advantage of this framework is its modularity: new datasets can be included with ease by adding ``measurement'' classes which converts any supported generator's output to a cross-section and compares it to data, without requiring the user to understand the output formats of the generators. Similarly, new generators can be added without requiring detailed understanding of the entire NUISANCE framework. The only dependency of NUISANCE outside the chosen generator(s) themselves is the ROOT library~\cite{Brun:1997pa}.

In this paper, we describe the core structure of NUISANCE, give the scope of the supported features, and demonstrate different usage scenarios. Detailed documentation of included datasets, validation plots, and usage instructions with examples can be found at \url{nuisance.hepforge.org}.

\section{NUISANCE}
\label{sec:code}
This section gives an overview of the core structure and design principles behind the NUISANCE framework. Full support for the standard output of the GENIE
\cite{Andreopoulos:2009rq,Andreopoulos:2015wxa}, 
NEUT \cite{Hayato:2009zz}, NuWro \cite{Juszczak:2009qa} and GiBUU~\cite{Buss:2011mx}
neutrino event generators is provided, with limited support for NUANCE\footnote{Only shape comparisons are possible with NUANCE because of limitations in the generator output.} \cite{Casper:2002sd}.
The core structure is designed to be easily extended, with
support for different event generators possible in later versions (e.g. neutrino, electron and pion scattering simulations).

\subsection{Input handling}
\begin{figure}
  \centering
  \begin{subfigure}{0.49\columnwidth}
    \includegraphics[width=\textwidth]{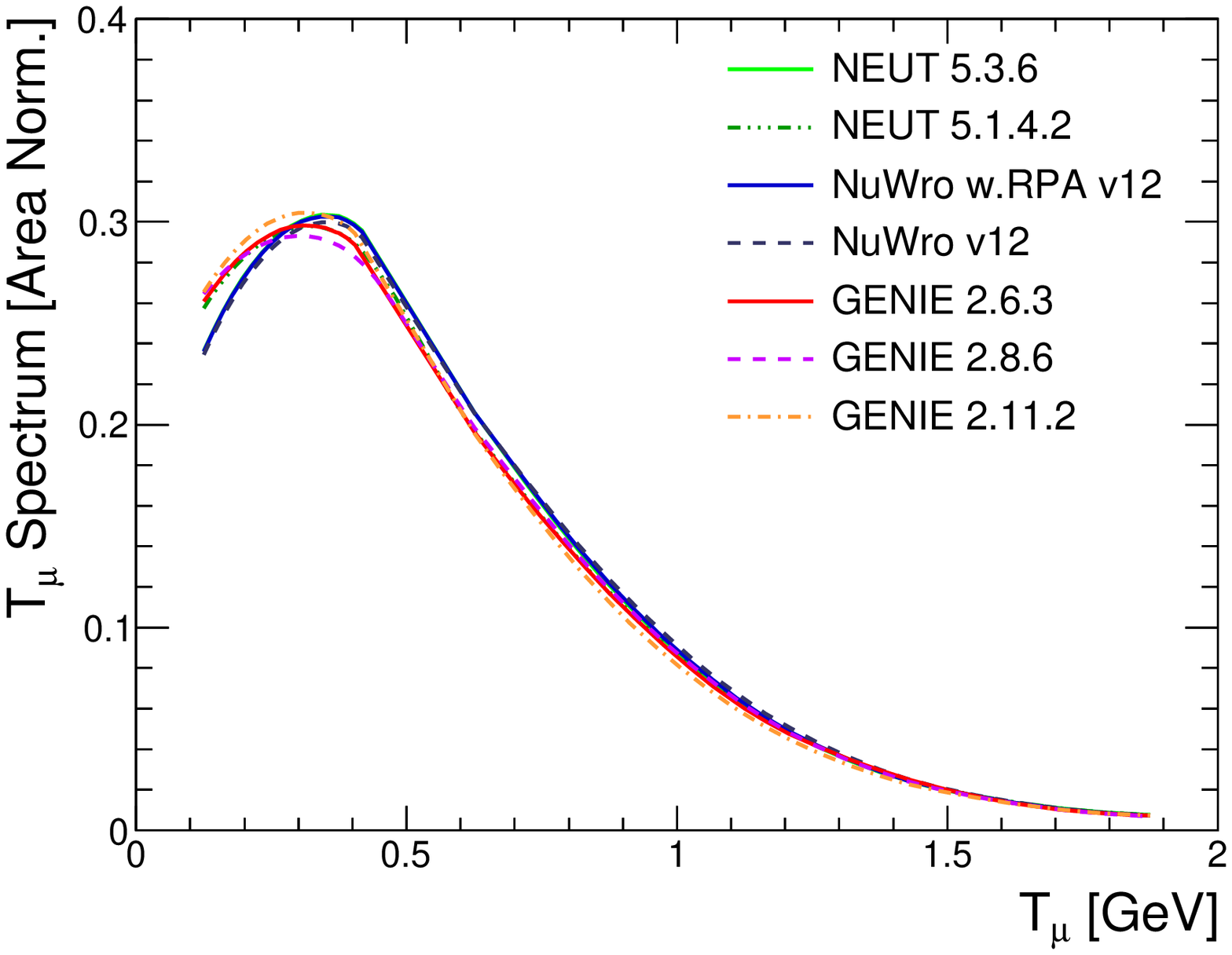}
    \caption{Muon kinetic energy}
  \end{subfigure}
  \begin{subfigure}{0.49\columnwidth}
    \includegraphics[width=\textwidth]{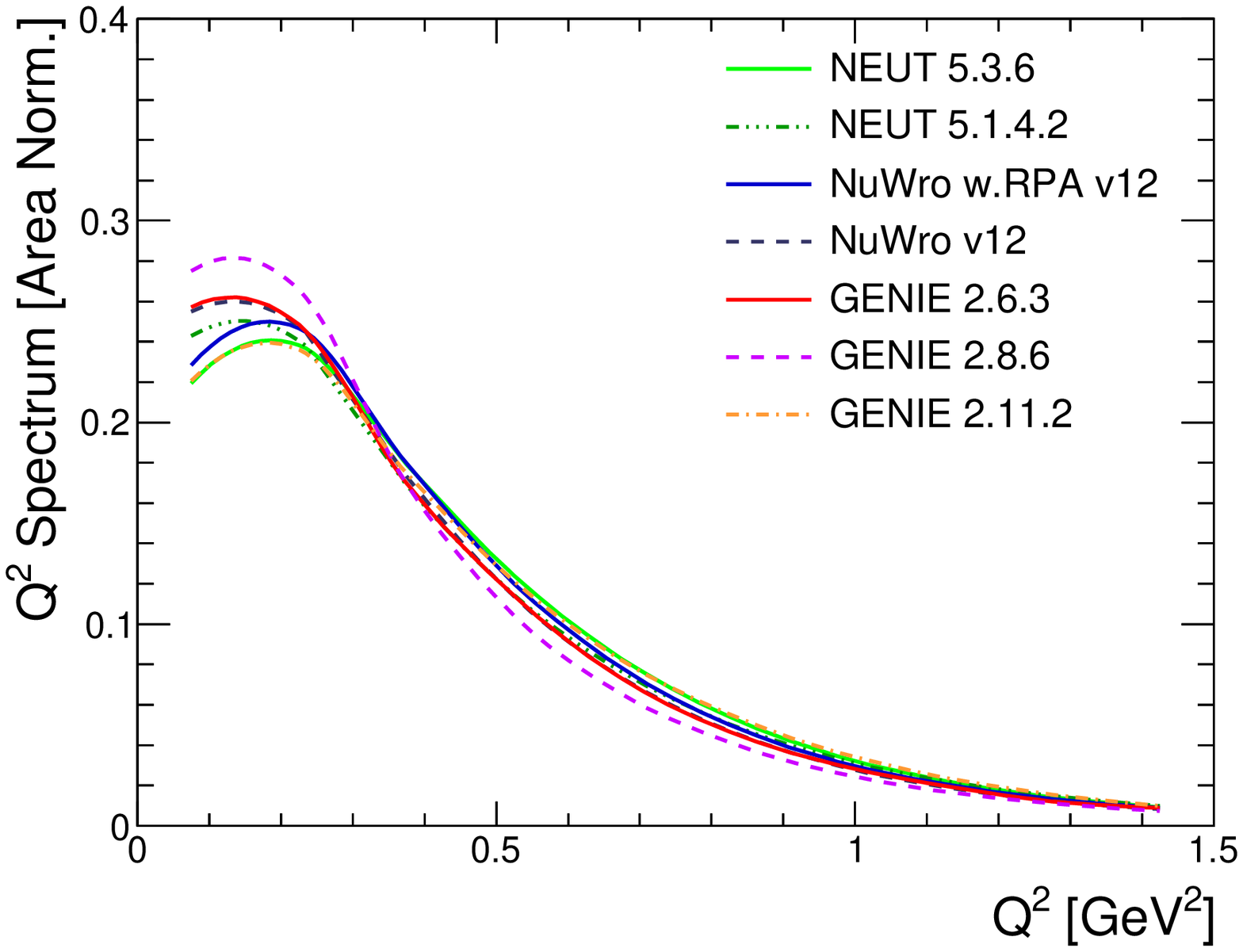}
    \caption{Four-momentum transfer}
  \end{subfigure}
  \caption{Shape comparison of CC-inclusive $\nu_{\mu}$--CH$_{2}$ events generated in different versions of the NEUT, NuWro, and
GENIE generators using the MiniBooNE neutrino-mode flux~\cite{mb-flux} (shown in Figure~\ref{fig:fluxes}).}
\label{fig:difxseccomp}
\end{figure}

\begin{figure}
\centering
\begin{subfigure}{0.49\columnwidth}
  \includegraphics[width=\textwidth]{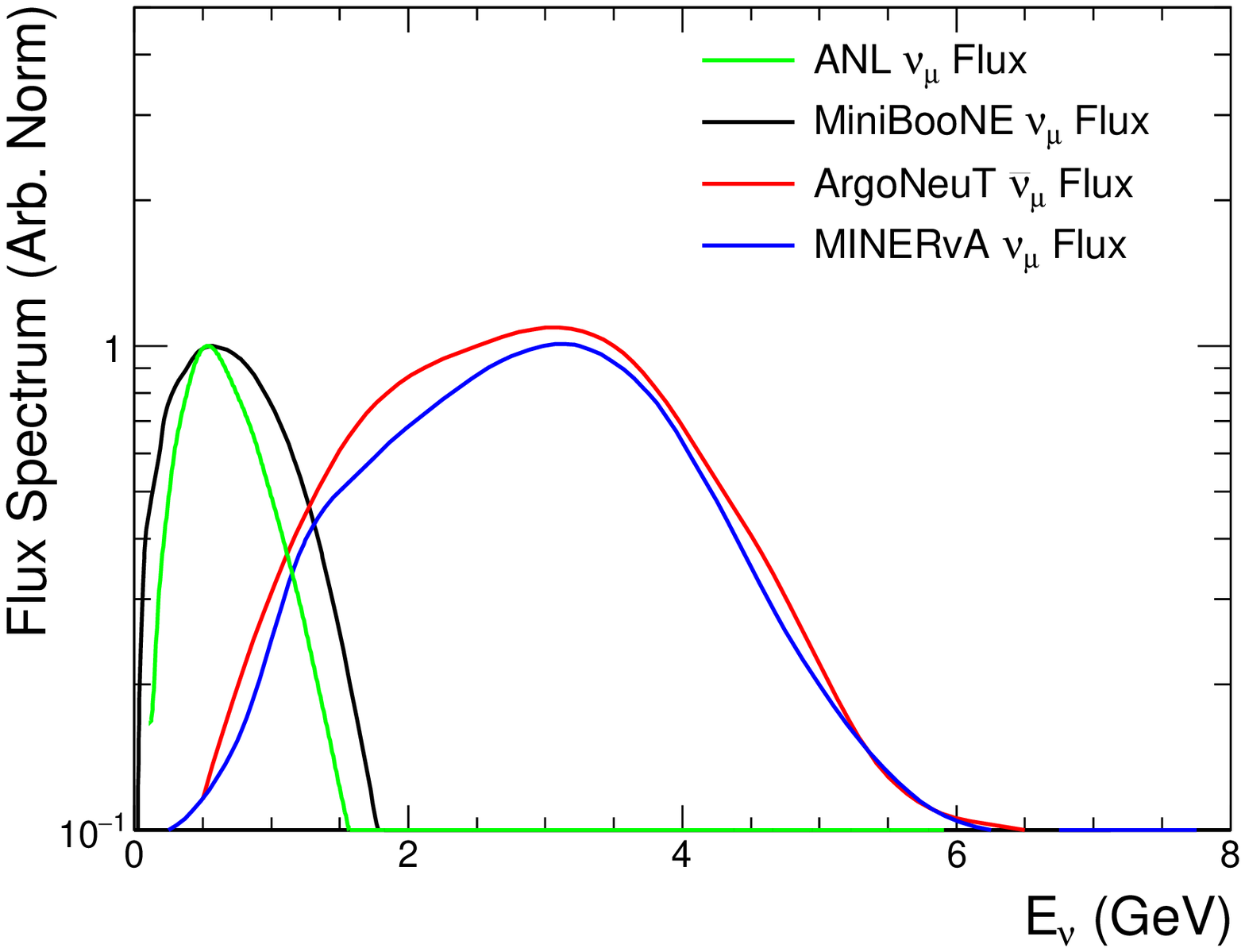}
  \caption{Linear energy scale}
\end{subfigure}
\begin{subfigure}{0.49\columnwidth}
  \includegraphics[width=\textwidth]{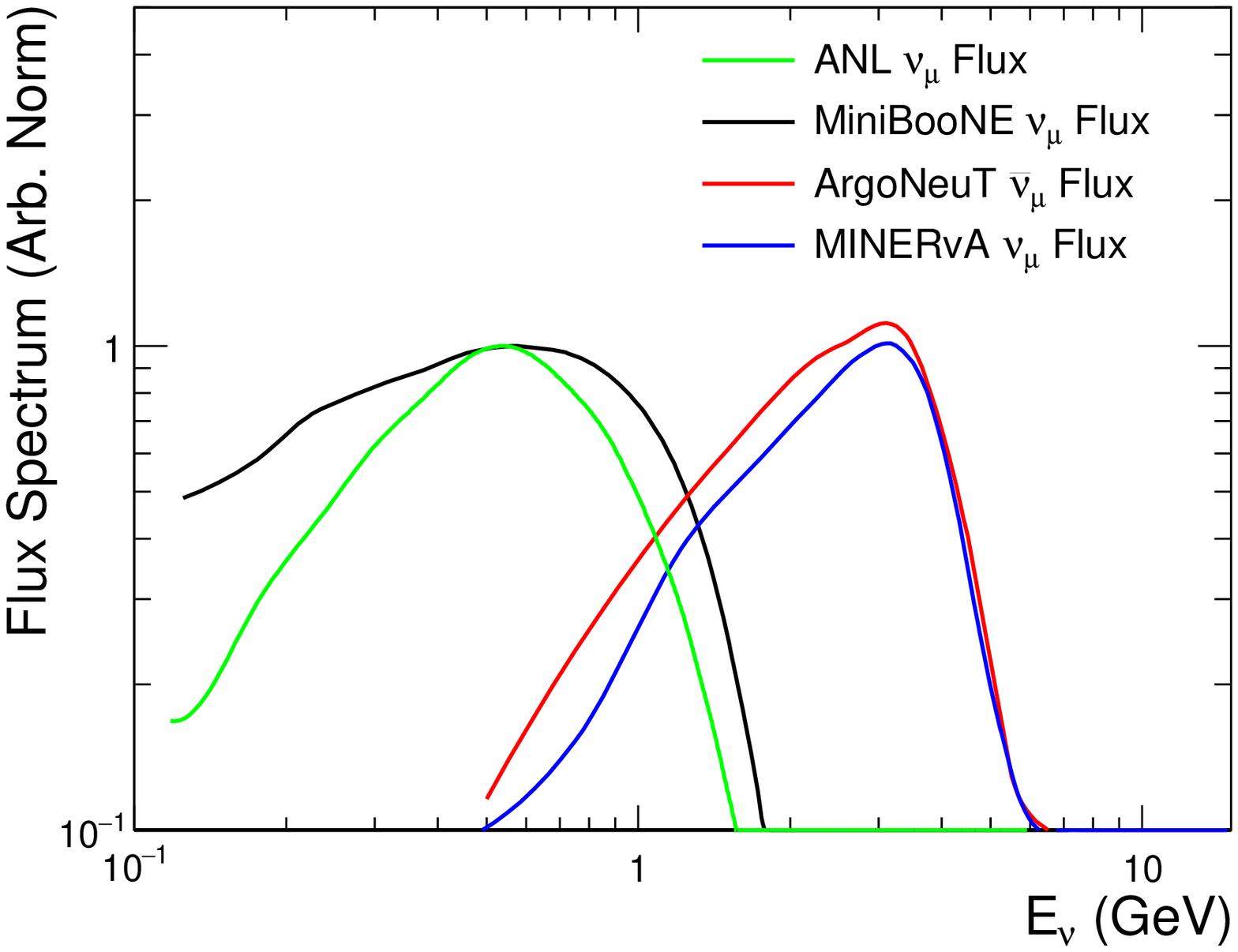}
  \caption{Log energy scale}
\end{subfigure}
  \caption{\label{fig:fluxes} Flux distributions used to generated events for NUISANCE comparisons in this work \cite{Barish:1977qk,mbCCQE,Eberly:2014mra,Acciarri:2014isz}, also supplied with NUISANCE.}
\end{figure}

Each event generator has a different output format and event structure,
but the underlying content is the same, always including:
\begin{itemize}
\item a list of incoming/outgoing particles, with their four-momentum, PDG code, and status;
\item an underlying interaction mode used in the generation\footnote{The definition of these channels may vary between generators. Since most measurements are topology based, this is not a problem.},
  e.g. $\bar{\nu}_{\mu}$--$^{12}$C CCQE, $\nu_{e}$--$^{16}$O CC1$\pi^{+}$;
\item a method for normalising the event distribution to produce a differential cross section;
\item (optional) information to support event reweighting, described in Section~\ref{sec:reweight}.
\end{itemize}
To ensure consistency between generators, and to increase speed, NUISANCE uses a reduced event structure that contains only the information required and unifies the
format for the various generators. The conversion to the standardised format is performed by the \emph{InputHandler} class when an event is first used and all subsequent NUISANCE functionality uses this format. The structure provides access to information about the event using common caller functions which are unified for the generators. It also ensures compatibility were new generators to be added to the \emph{InputHandler} in the future. Figure~\ref{fig:difxseccomp} illustrates a simple comparison of CC-inclusive $\nu_{\mu}$--CH$_{2}$ events generated with a variety of generators using the MiniBooNE neutrino-mode flux shown in Figure~\ref{fig:fluxes}~\cite{mb-flux}.

\subsection{Cross section normalisation}
The NUISANCE \emph{InputHandler} calculates all the information needed to weight events correctly such that the final distribution is normalised to an inclusive cross-section prediction. In the case of GiBUU, these weights are calculated by the generator itself and saved with the event. For the other generators---where the number of events from different interaction channels are generated in proportion to their cross section---a single weight is calculated which is applied to all events. NUISANCE requires the flux distribution used to generate the sample, $\phi(E_{\nu})$, and calculates the predicted total event rate per target nucleon, $R$, from information available in the generator output file,
\begin{equation}
  R = \int \sigma_{tot}(E_{\nu}) \times \phi(E_{\nu}) dE_{\nu}, 
\end{equation}
\noindent where $\sigma_{tot}(E_{\nu})$ is the \emph{total inclusive cross section} as a function of neutrino energy and the integral runs over the entire energy range the events were generated in\footnote{Even if the signal definition contains a cut on $E_\nu$}. $R$ is provided in the default output of the NuWro and NEUT generators, but must be calculated for GENIE from the event record. A separate application, \emph{PrepareGENIE}, is supplied to reconstruct the predicted GENIE cross section as a function of neutrino energy for each interaction channel. These cross-section ``splines'' are then used to predict $R$ for the event sample given the input flux.

\begin{figure}
 \begin{subfigure}{0.49\columnwidth}
    \includegraphics[width=\textwidth]{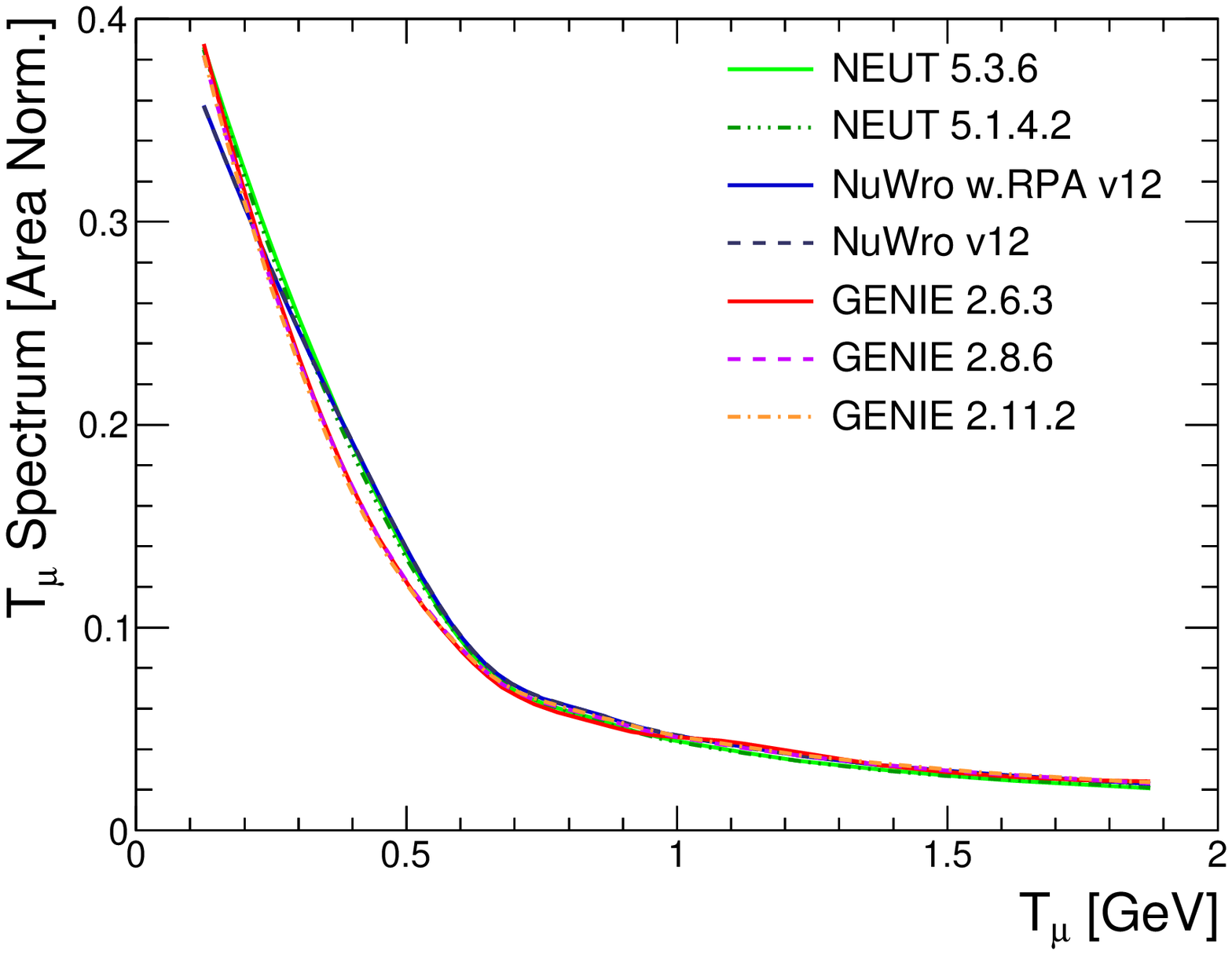}
    \caption{Raw event rate}
  \end{subfigure}
  \begin{subfigure}{0.49\columnwidth}
    \includegraphics[width=\textwidth]{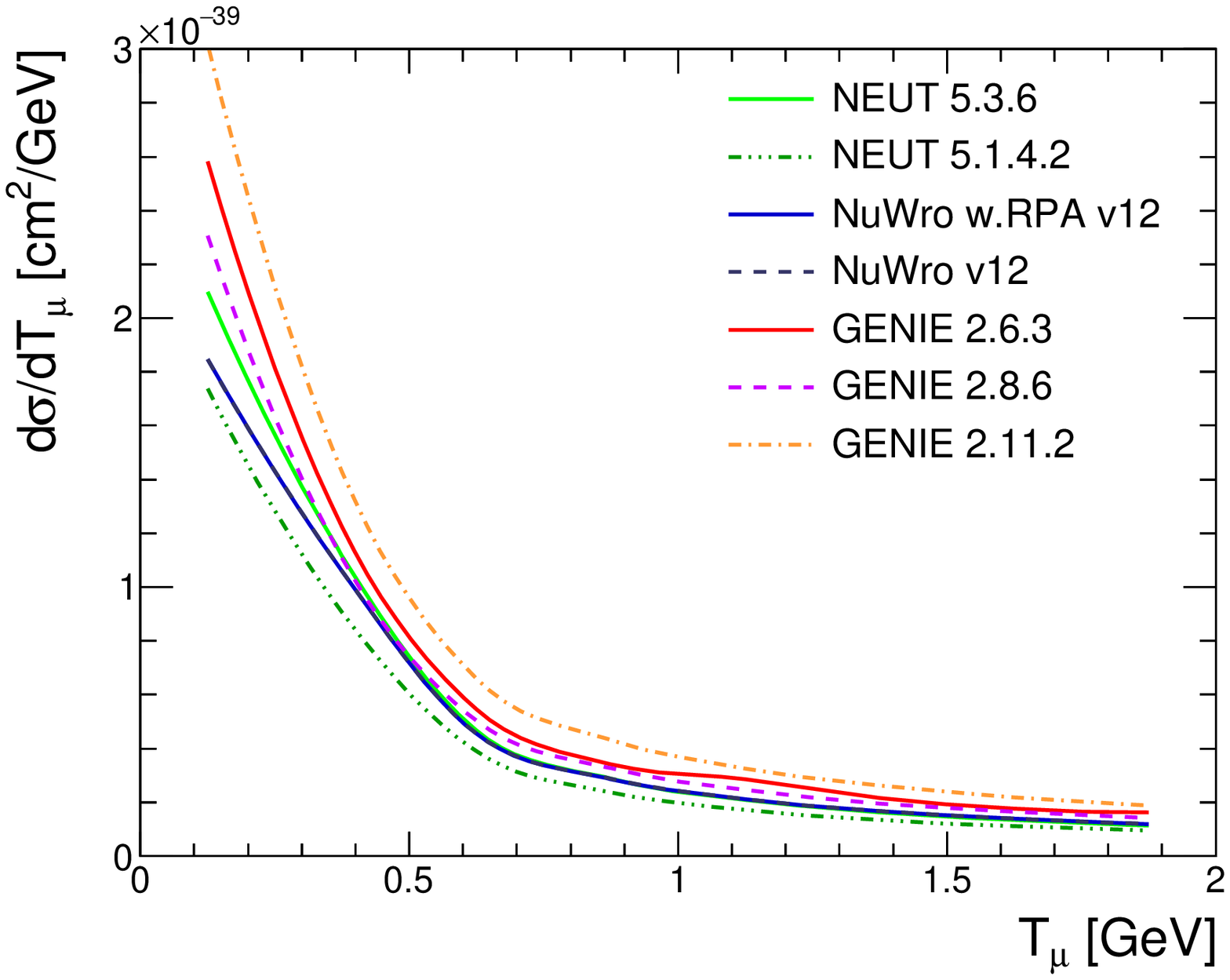}
    \caption{Cross section}
  \end{subfigure}
  \caption{Raw event rate and correctly normalised cross-section distributions as a function of muon kinetic energy, $T_{\mu}$, shown for
    $\nu_{\mu}$--CH$_{2}$ CC1$\pi^{\pm}$ events produced using a variety of generators and the MiniBooNE neutrino-mode flux (shown in Figure~\ref{fig:fluxes}).}
  \label{fig:xsec_norm}
\end{figure}

A final \emph{flux-averaged} cross-section weight, $W$, can then be calculated for each generator 
\begin{equation}
  W = \frac{ R }{N \Phi},
  \label{eq:xsec_weight}
\end{equation}
\noindent where $N$ is the total number of events generated in the generator, and $\Phi$ is the integrated neutrino flux between $E_\nu^{min}$ and $E_\nu^{max}$ (the neutrino energy limits in the signal definition)
\begin{equation}
\Phi = \int_{E_\nu^{min}}^{E_\nu^{max}} \phi(E_\nu) dE_\nu.
\label{eq:flux_int}
\end{equation}
Filling a histogram in interaction variable $x$ with the weights $W$, for events that pass a user-supplied signal definition produces a correctly normalised \emph{flux-averaged} cross section $d\sigma(x)$. Dividing by each bin's width produces the differential cross section $d\sigma(x)/dx$, shown in Figure~\ref{fig:xsec_norm}. This is only appropriate when comparing to \emph{flux-averaged} cross-section results. For a \emph{flux-integrated} cross section the flux is instead integrated out on a bin-by-bin basis and the $E_\nu^{min},E_\nu^{max}$ limits in \ref{eq:flux_int} are instead given by the bin-edges of the relevant $E_{\nu}$ bin that an event resides in.

\subsection{Comparisons to cross-section data}
To compare different models to a chosen neutrino cross-section dataset a model prediction must be produced
that matches the original data analysis, matching true signal and binning definitions. To add a new dataset to NUISANCE a new ``measurement'' class must be created which defines the analysis
method used to turn a set of generated MC events into a matching model prediction. To minimise the work required
for a user to add a new dataset, these classes inherit from a small number of ``measurement'' base classes which
define methods common for all cross-section predictions of a certain type (e.g. one-dimensional, two-dimensional).
Comparisons can be added into the framework provided the following information is known:
\begin{enumerate}
\item {\bfseries Data distribution:} the measured central values and uncertainties must be supplied in either text or ROOT file format.

\item {\bfseries Signal definition:} a signal which acts to select events using the particle list must be defined. Utility definitions are available for common signal definitions, e.g. CC$0\pi$.

\item {\bfseries Binning definition:} the kinematic variables to plot must be defined from incoming/outgoing particle list, e.g. lepton momentum. The binning itself is copied from the data distribution.

\item {\bfseries (Optional) Covariance:} the correlations between each point in the data distribution so that a more accurate likelihood can be formed. If no covariance is provided uncorrelated errors are assumed on each point.
  \item {\bfseries (Optional) Smearing matrix:} a translation matrix to smear true variable distributions, converting them into detector reconstructed variable distributions that can be compared directly to reconstructed data releases.
\end{enumerate}
Every sample has an event loop in the base class which iterates over all events given an input file provided at runtime. The input file to NUISANCE is the output of the generator(s) of interest\footnote{Which has to be pre-processed when using GENIE with the \emph{PrepareGENIE} utility.}. Only events which pass the signal definition are retained past the first event loop. For signal events, a cross-section weight is calculated using Equation~\ref{eq:xsec_weight} (assuming a flux-averaged cross section), and all histograms in the specific measurement class are filled. This automated event loop ensures that the core handling of event inputs remains the same for every measurement implementation class, although each method can be overloaded if necessary.

An \emph{event manager} can be turned on to avoid iterating over events in the same input file multiple times if two or more measurement classes use the same generator output file. When it is used, the event manager checks whether the signal criteria are met in any of the classes, and retains events which are signal for one or more of them. This can significantly speed up NUISANCE for many of the fitting routines, in which weights need to be recalculated and histograms refilled multiple times for a number of datasets, e.g. comparing multiple kinematic distributions of the same measurement, or different measurements using the same flux.

The most basic measurement implementation class produces a correctly normalised histogram with the same binning as the data and can be compared directly. ROOT histograms showing the data, MC prediction(s) and the input flux(es) are saved in the output file for later comparison. Various utility functions exist to include histograms, e.g. stacking the MC prediction by interaction mode or particle type, shape predictions and data--MC ratios, as in \autoref{fig:additionaloutput}. It is a trivial exercise to include any additional histograms by overloading the base-class functions.

The data--MC agreement is evaluated by a likelihood which is saved in the output file. The base class defaults to using a covariance matrix if supplied, or reverts to a Gaussian pdf for cross-section measurements and a Poisson pdf for event-rate measurements.
\begin{align}
  \mbox{(Covariance)~~} -2LL &=
  \sum_{ij} (\nu^{\mtlabel[data]}_{i} - \nu^{\mtlabel[MC]}_{i}) (M^{-1})_{ij} (\nu^{\mtlabel[data]}_{j} - \nu^{\mtlabel[MC]}_{j}) \\
  \mbox{(Gaussian pdf)~~} -2LL &=
  \sum_{i} \left( \frac{\nu^{\mtlabel[data]}_{i} - \nu^{\mtlabel[MC]}_{i}}{\sigma^{\mtlabel[data]}_{i}} \right)^{2} \\
  \mbox{(Possion pdf)~~} -2LL &=
  2\sum_{i} \nu^{\mtlabel[MC]}_{i} - \nu^{\mtlabel[data]}_{i} + \nu^{\mtlabel[data]}_{i} log\left( \frac{\nu^{\mtlabel[data]}_{i}}{\nu^{\mtlabel[MC]}_{i}}\right)
\end{align}
where $\nu_{i}$ is the bin content in $i$-th bin for data or MC, $M$ is the supplied covariance matrix, and $\sigma_i$ is the error on the $i$-th bin in data.

These likelihood functions can easily be overloaded by the user for each measurement class to allow more complex likelihoods to be used for a given analysis such as shape-only and floating normalisation likelihoods.



\subsection{Event reweighting}
\label{sec:reweight}
Event reweighting allows MC predictions to be modified after event generation by separating out parts of the cross section that can be recalculated without having to perform the entire MC simulation again. This saves considerable computation time and is useful for both model tuning and the evaluation of model systematic uncertainties since the events are already generated.

NUISANCE has native support for the NEUT, NuWro~\cite{Pickering:2016icq}, and GENIE event reweighting libraries.
A NUISANCE reweight wrapper is provided which can read in the custom event format and return an event weight for any given parameter set. All reweightable parameters provided by the external libraries can be used to calculate new event weights. This allows the user to easily modify the NUISANCE prediction in the various fitting and validation routines, whilst requiring only minimal knowledge of the individual generators reweight engines. The reweighting parameters are specified in the user-supplied card files.

\section{NUISANCE applications}
A number of different applications are available with the NUISANCE framework, all of which are controlled by card files. In this section, the main NUISANCE applications are introduced and a general overview of their functionality is given. Detailed information on the input format required and example card files are available at \url{nuisance.hepforge.org}. Additionally, the general behaviour of NUISANCE can be controlled by a configuration file, which can be overridden for specific comparisons if desired either in the card file or with command line arguments.

\subsection{Simple data--MC comparisons: \emph{nuiscomp}}
\label{sec:nuiscomp}
The simplest usage case for NUISANCE is to produce an MC prediction for one or multiple measurement(s). The \emph{nuiscomp} application accepts a simple card file with a list of datasets to produce comparisons for and the input file to generate them with. It saves the resulting histograms to a single ROOT output file. Optionally, the user can specify any reweightable parameters which should be set when making this comparison. The default behaviour is to not reweight the prediction.

\begin{figure}
  \centering
  \begin{subfigure}{0.49\columnwidth}
    \includegraphics[width=\textwidth]{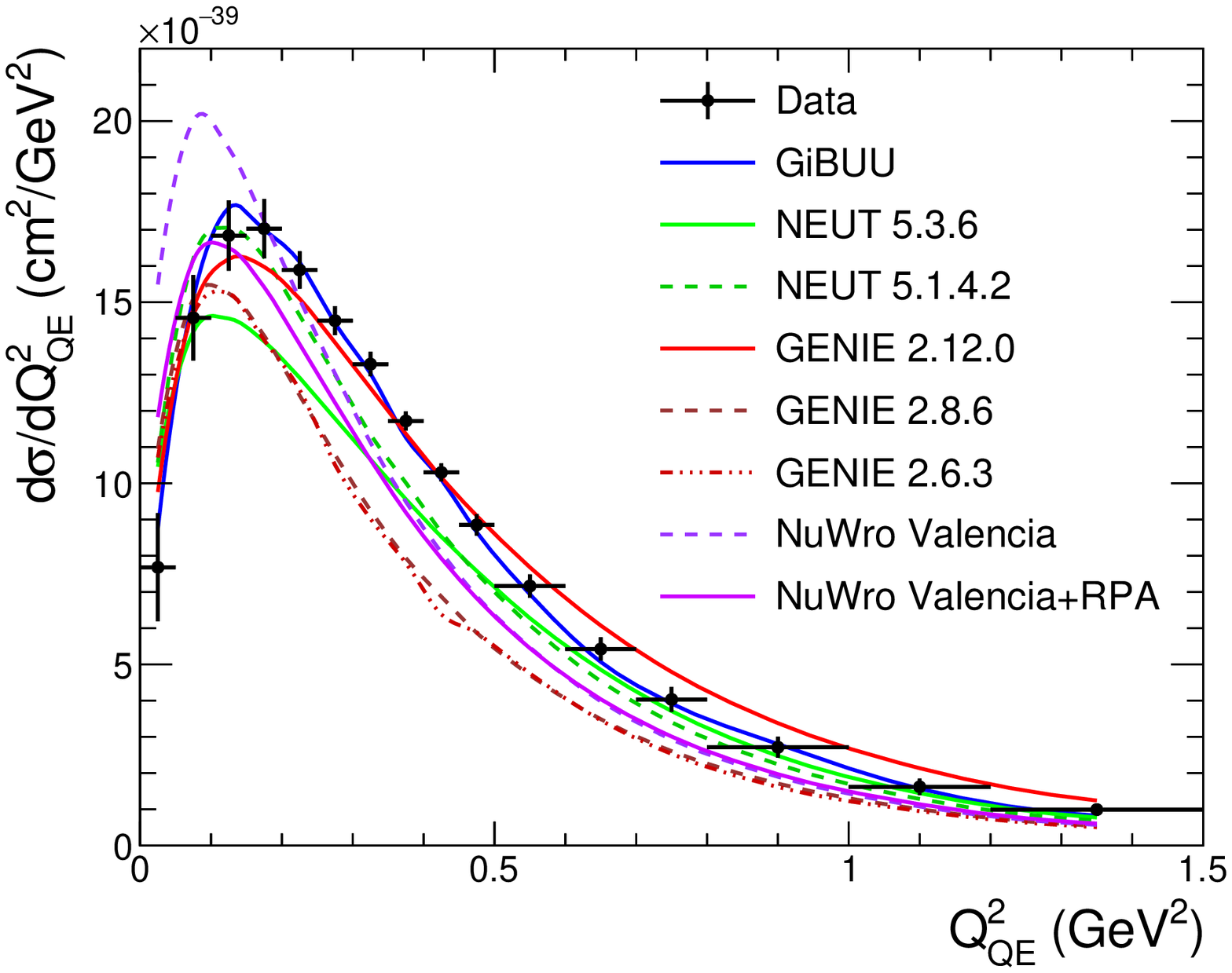}
    \caption{MiniBooNE $\nu$--CH$_{2}$ CCQE}
  \end{subfigure}
  \begin{subfigure}{0.49\columnwidth}
    \includegraphics[width=\textwidth]{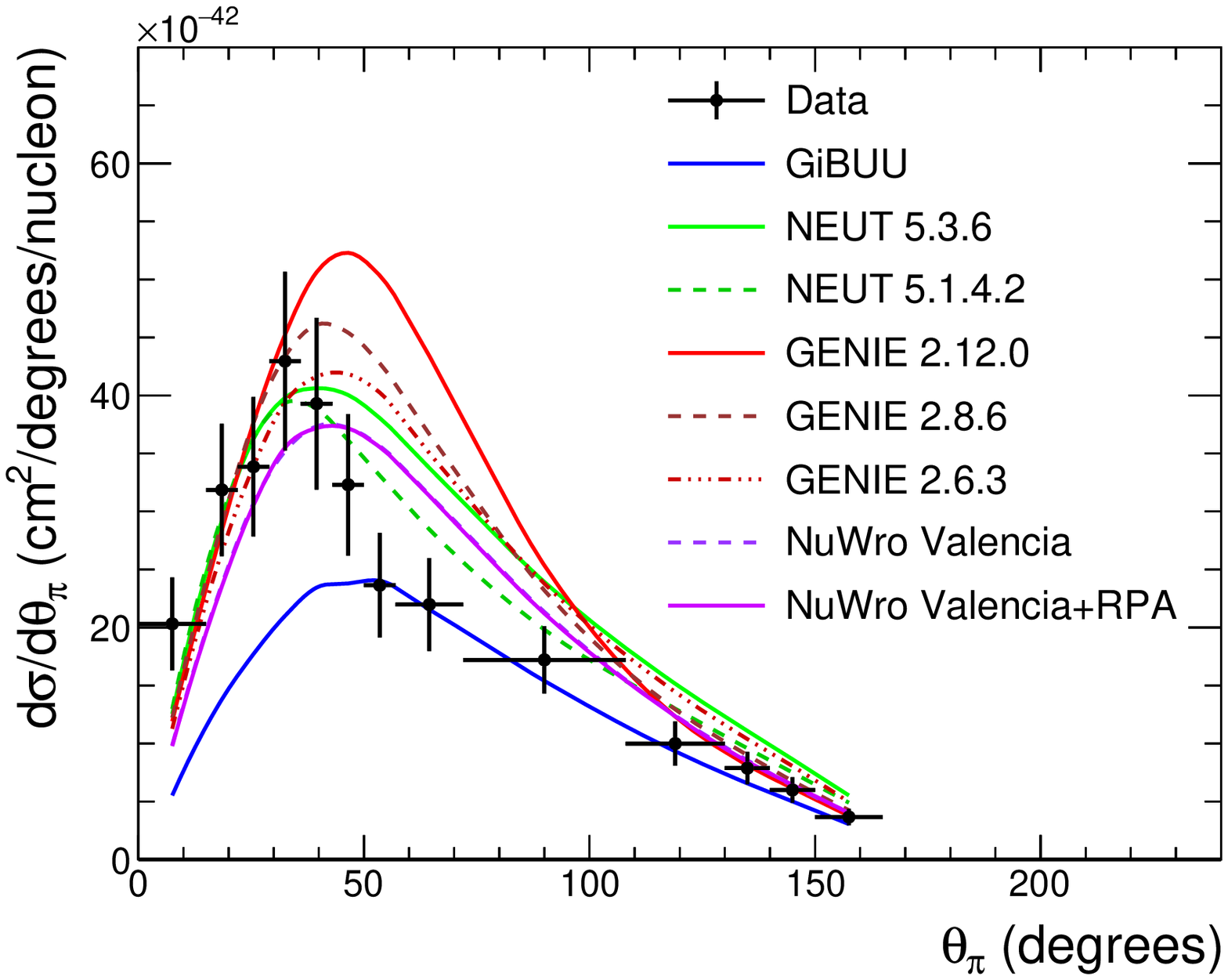}
    \caption{\minerva $\nu$--CH CC$1\pi^{\pm}$}
  \end{subfigure}
  \caption{Comparisons of different generators and their versions to published measurements~\cite{mbCCQE,Eberly:2014mra}. The fluxes used to produce these distributions are shown in Figure~\ref{fig:fluxes}.}
\label{fig:generatorcomp}
\end{figure}

As described in Section~\ref{sec:code}, different input types are automatically handled by NUISANCE. Changing generators only requires changing the input type and the file location, both specified by the user in the card file. Multiple generator, multiple dataset comparisons are straightforward with NUISANCE, as illustrated in \autoref{fig:generatorcomp}.

\begin{figure}
  \begin{subfigure}{0.49\columnwidth}
    \includegraphics[width=\textwidth]{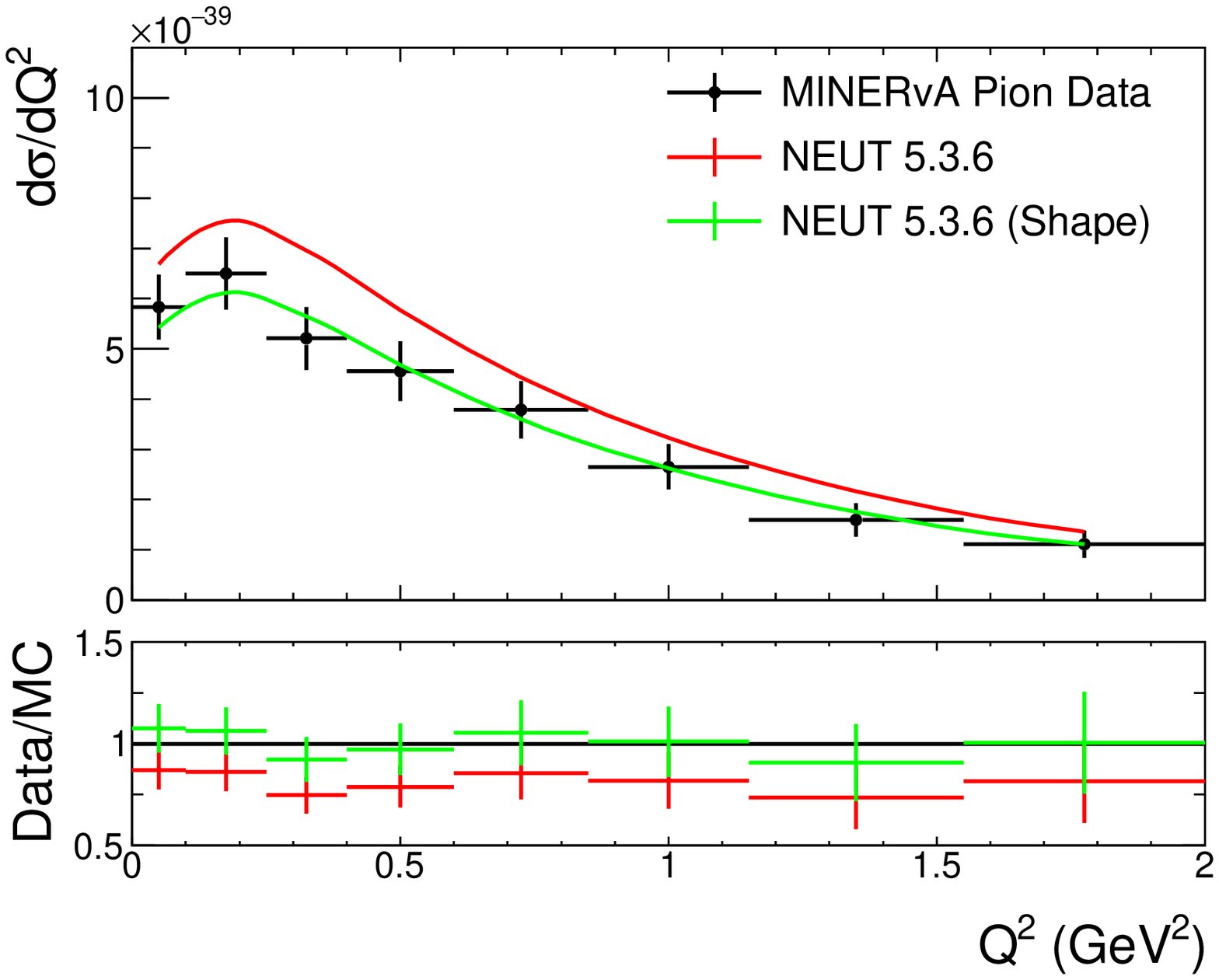}
    \caption{Rate and shape-only comparisons to data}
  \end{subfigure}
  \begin{subfigure}{0.49\columnwidth}
    \includegraphics[width=\textwidth]{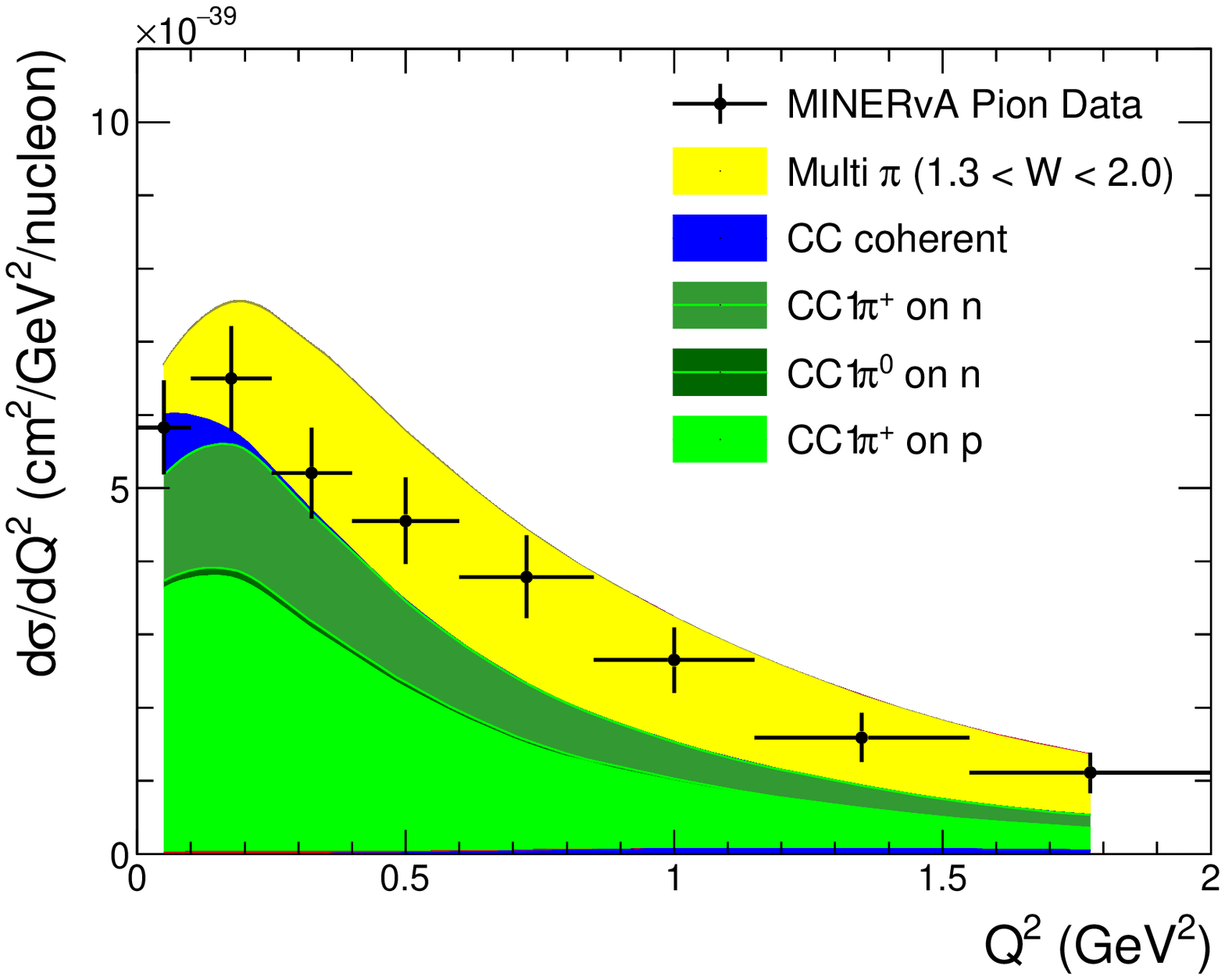}
    \caption{MC broken down by true interaction mode}
  \end{subfigure}
  \caption{NEUT prediction for \minerva CC$N\pi^+$~\cite{McGivern:2016bwh} produced with the \emph{nuiscomp} application. The fluxes used to produce these distributions are shown in Figure~\ref{fig:fluxes}.
\label{fig:additionaloutput}  }
\end{figure}

Alongside the data and MC histograms, the \emph{nuiscomp} application saves a number of auxiliary MC histograms to help evaluate where there are tensions between the data and models. Examples include predictions separated by true interaction modes, and shape-only comparisons, shown in \autoref{fig:additionaloutput}.

\subsection{Raw generator comparisons: \emph{nuisflat}}

The \emph{nuisflat} application is intended to compare generators to each other, rather than to data. A template class converts the generator events into a flat ROOT tree containing particle information for each event, what signal definitions the event passes, its interaction mode, amongst a host of other event variables. Additional quantities can be added to the tree for tailored studies including an option to save the entire NUISANCE event for access to the full input/output particle stack. Once the output has been produce by \emph{nuisflat} it loses all dependencies on the generator libraries. Analysis is only dependent on ROOT to inspect the tree and its contents; the application produces consistent generator comparisons with minimal knowledge of the individual generator.

As with all NUISANCE applications, \emph{nuisflat} supports reweighting the generated events to parameter variations specified in the input card file by the user at runtime.

\subsection{Systematic validation studies: \emph{nuissyst}}
\begin{figure}
\centering
\begin{subfigure}{0.49\columnwidth}
    \includegraphics[width=\textwidth]{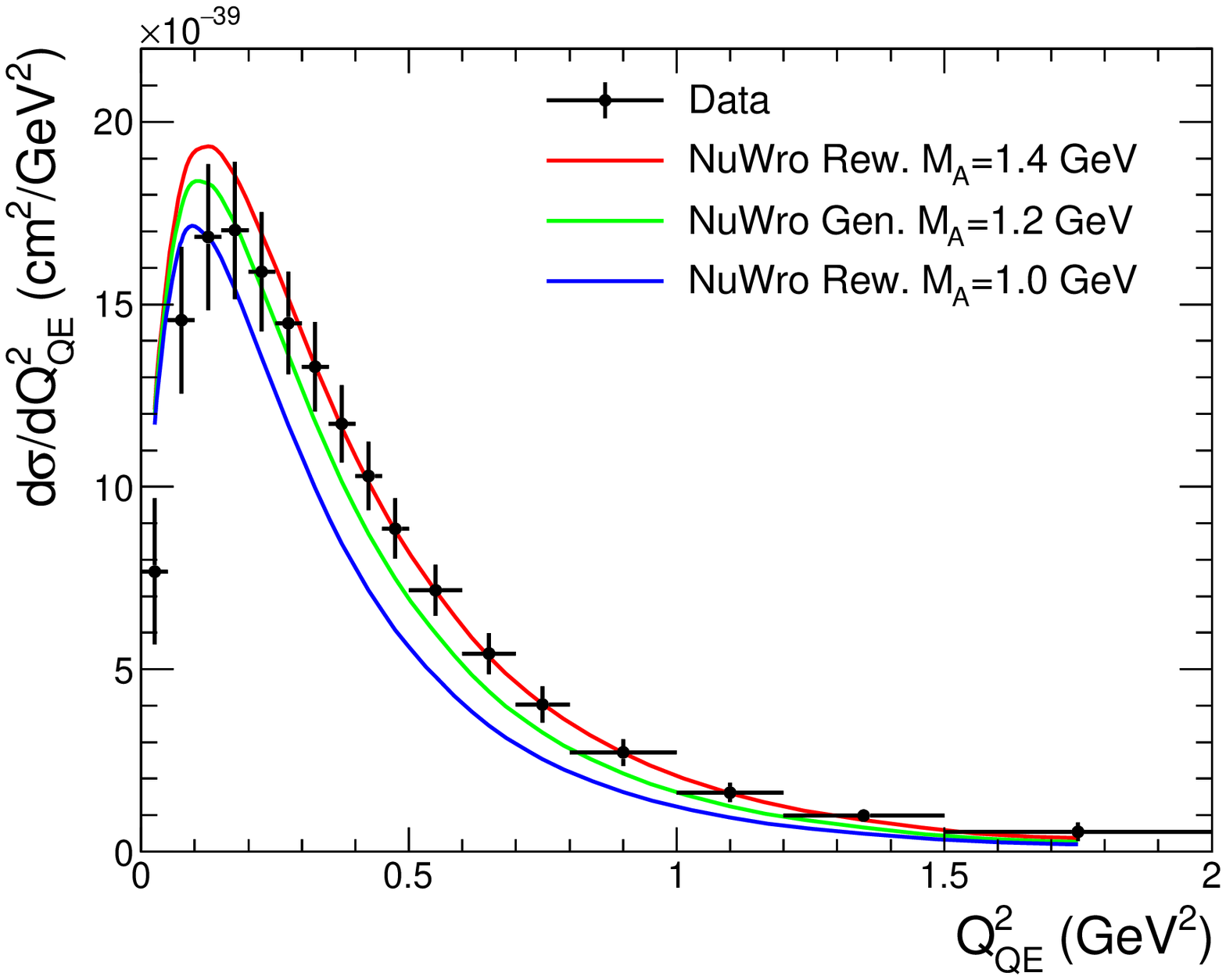}
    \caption{$M_{\mathrm{A}}$ variations}
  \end{subfigure}
  \begin{subfigure}{0.49\columnwidth}
    \includegraphics[width=\textwidth]{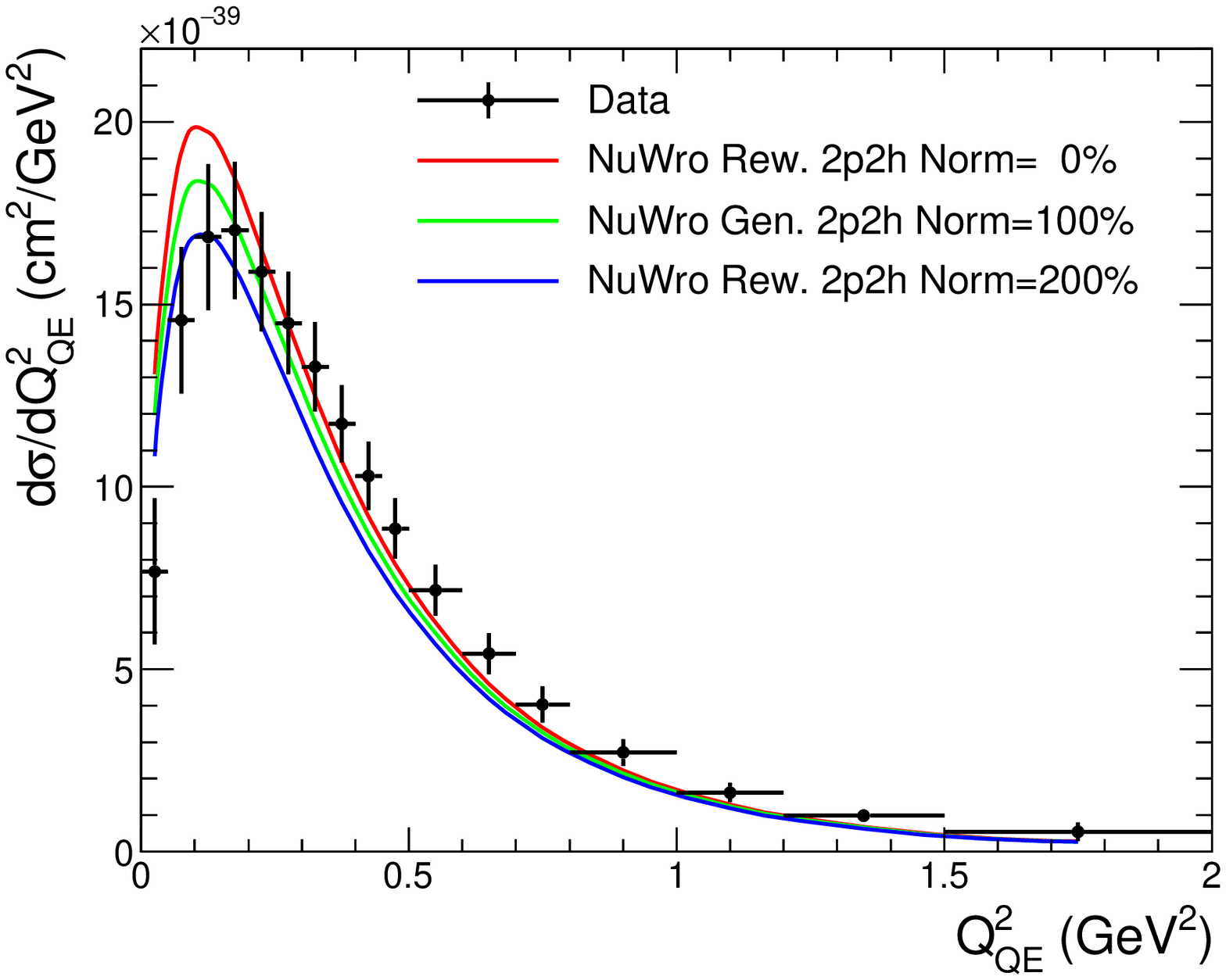}
    \caption{2p2h variations}
  \end{subfigure}
\caption{Reweighting responses for a set of NuWro events compared to MiniBooNE CCQE neutrino data.}
\label{fig:rwresponse}
\end{figure}

The \emph{nuissyst} application can be used to study the effects of cross-section systematics on user-specified distributions in a number of ways, provided that the generators have reweighting libraries enabled\footnote{Possible in GENIE, NEUT and NuWro}. It can step through a range of values for a reweightable parameter\footnote{If a parameter is reweightable by a generator's reweighting library it is supported in NUISANCE.} and validate reweighting implementations. It can also compare each generator's implementation of the reweighting engines---e.g. the effect of varying $M_{\mathrm{A}}^{\mbox{CCQE}}$ by $0.1\sigma$ in GENIE versus the same variation in NuWro. Examples of the output of this utility can be found in Figure~\ref{fig:rwresponse}.

\begin{figure}
\centering
  \begin{subfigure}{0.49\columnwidth}
    \includegraphics[width=\textwidth]{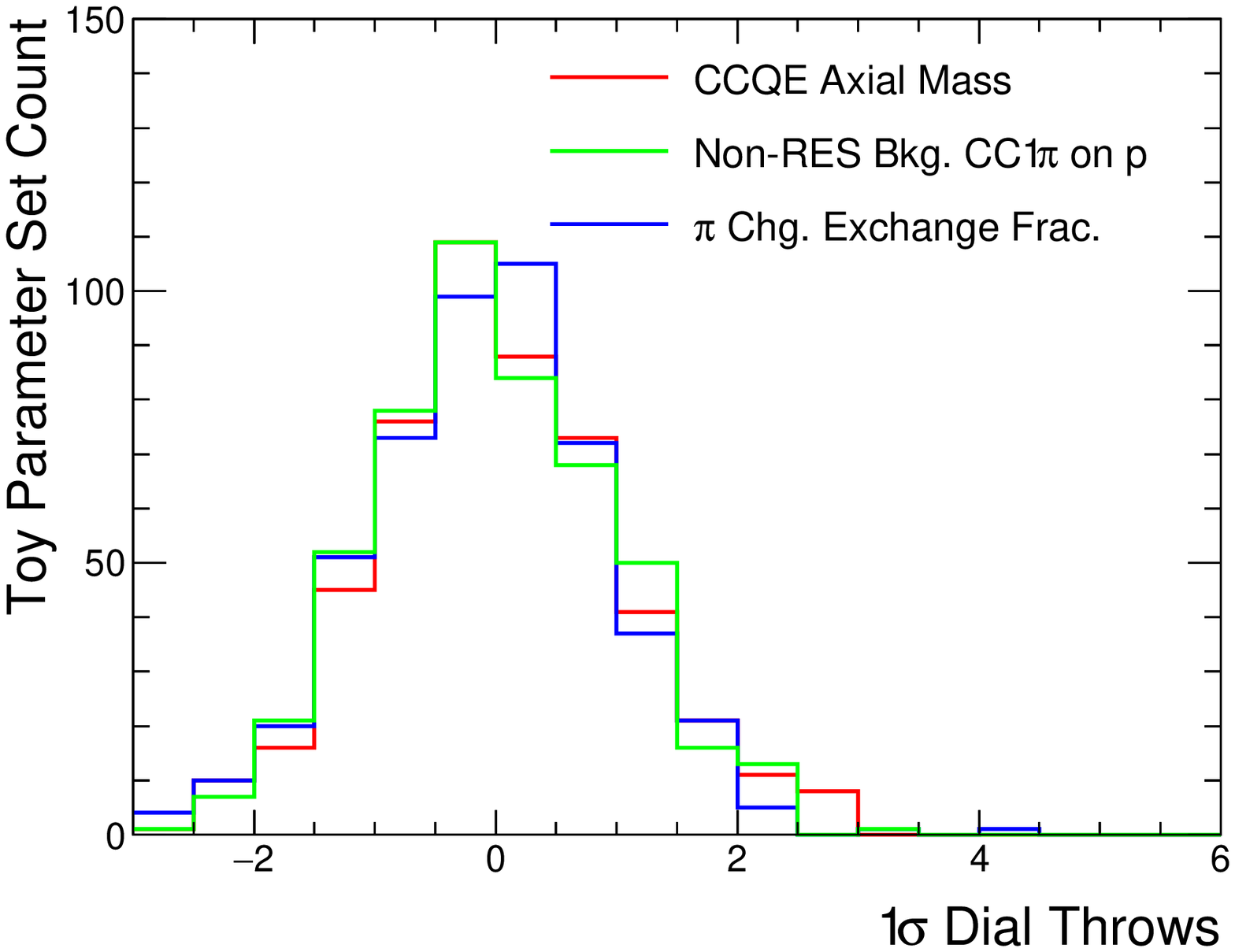}
    \caption{Thrown parameter values}
  \end{subfigure}
  \begin{subfigure}{0.49\columnwidth}
    \includegraphics[width=\textwidth]{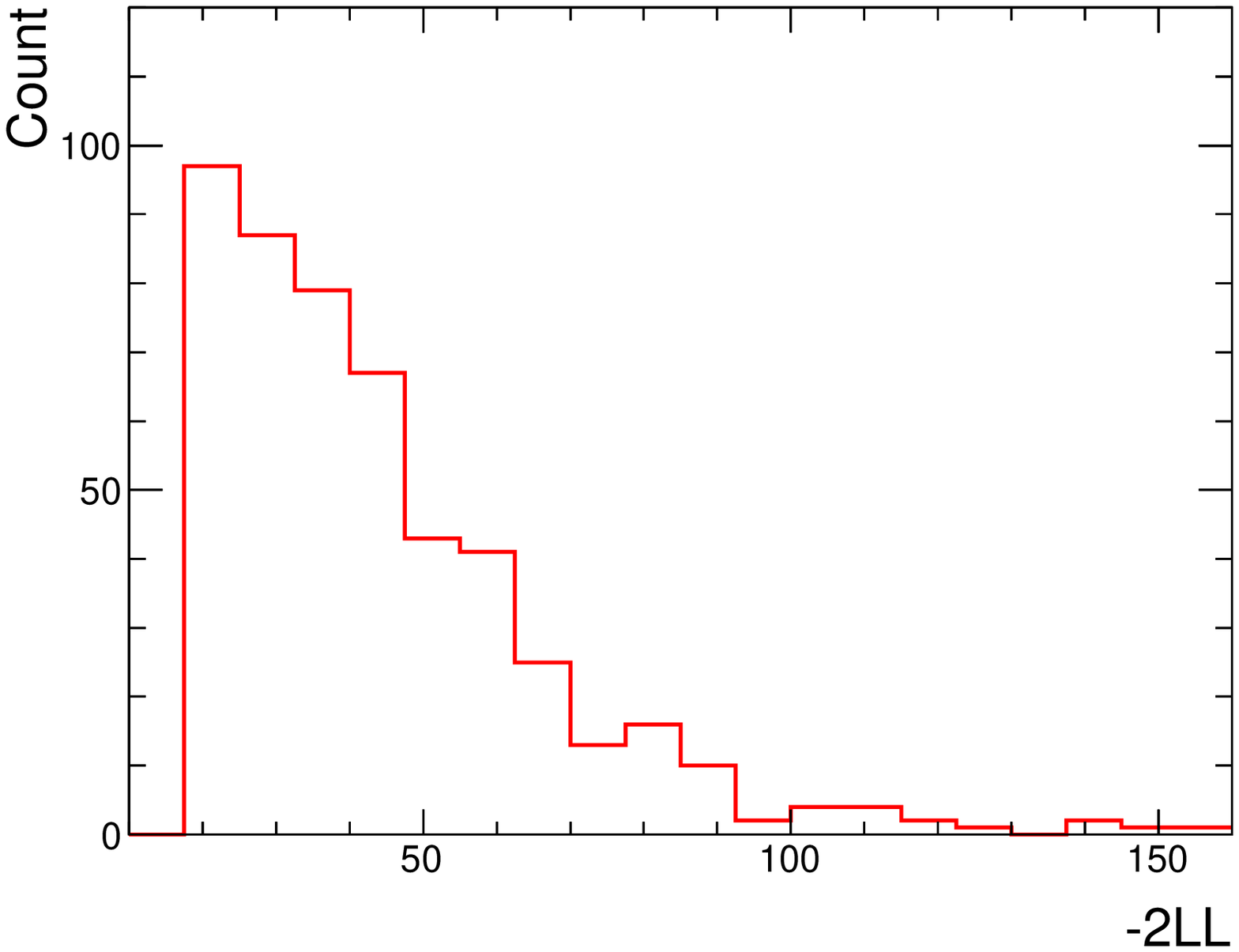}
    \caption{Throws Likelihood distribution}
  \end{subfigure}
\caption{Examples of thrown parameter values and the likelihood results for 500 different toy datasets generated with GENIE v2.12.0 for comparison to the ArgoNeut $\bar{\nu}$--$^{40}$Ar CC-inclusive dataset~\cite{Acciarri:2014isz}.
  \label{fig:toymcthrows}
  }
\end{figure}

The \emph{nuissyst} application can also make throws of any number of reweightable parameters to build up an error band on a generator prediction across any combination of datasets. The central value and $1\sigma$ uncertainty can be defined by the user, and Gaussian throws will be made around that value with the defined width. Flat parameter throws are also supported, and a method is included for throwing parameters according to a user-supplied parameter covariance matrix. In the latter case the distribution of parameter values for each bin of the requested distributions is used to produce a error band for that bin as shown in Figure~\ref{fig:toymcthrows}. 

\begin{figure}
\centering
  \begin{subfigure}{0.49\columnwidth}
    \includegraphics[width=\textwidth]{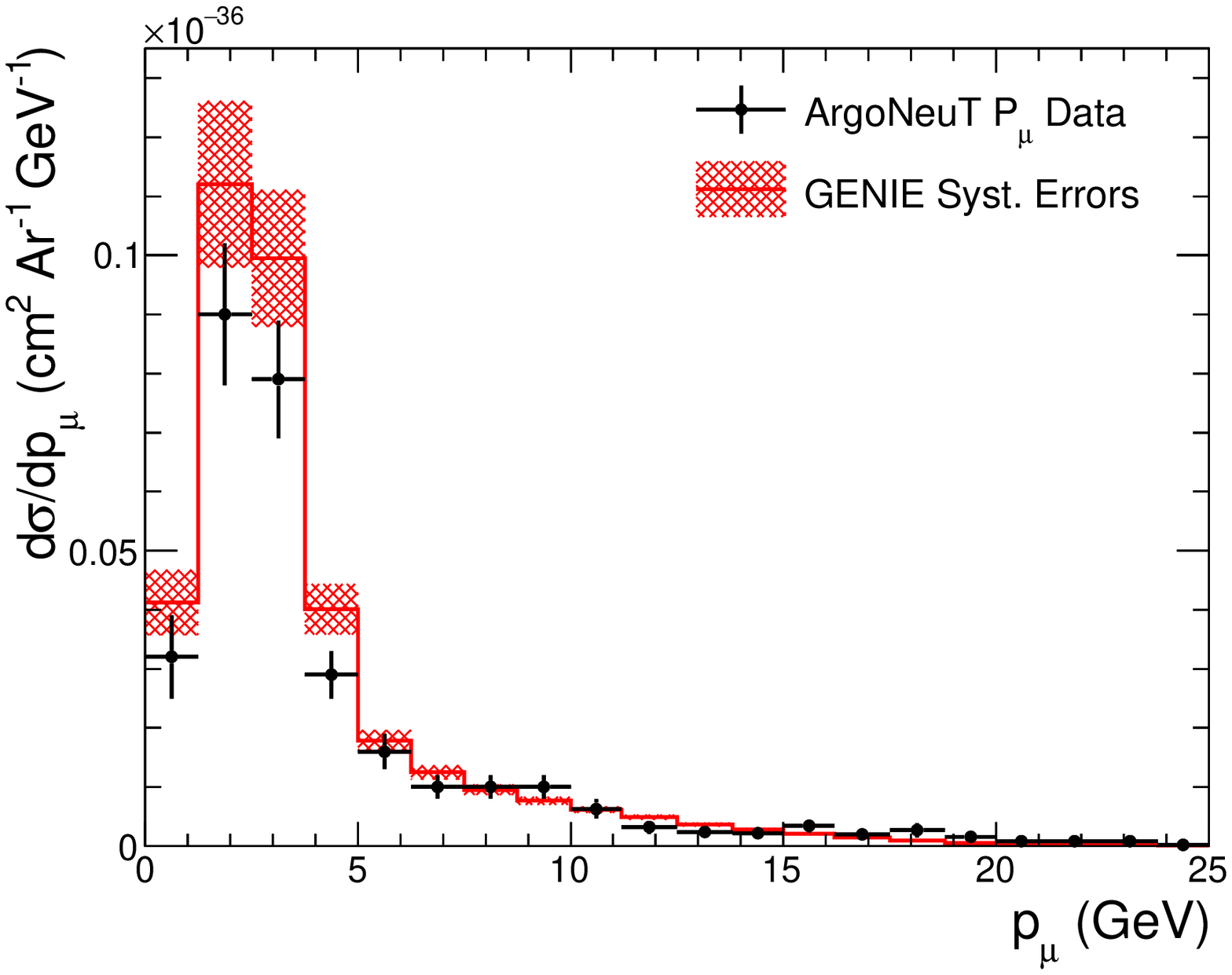}
    \caption{Muon momentum ($p_{\mu}$)}
  \end{subfigure}
  \begin{subfigure}{0.49\columnwidth}
    \includegraphics[width=\textwidth]{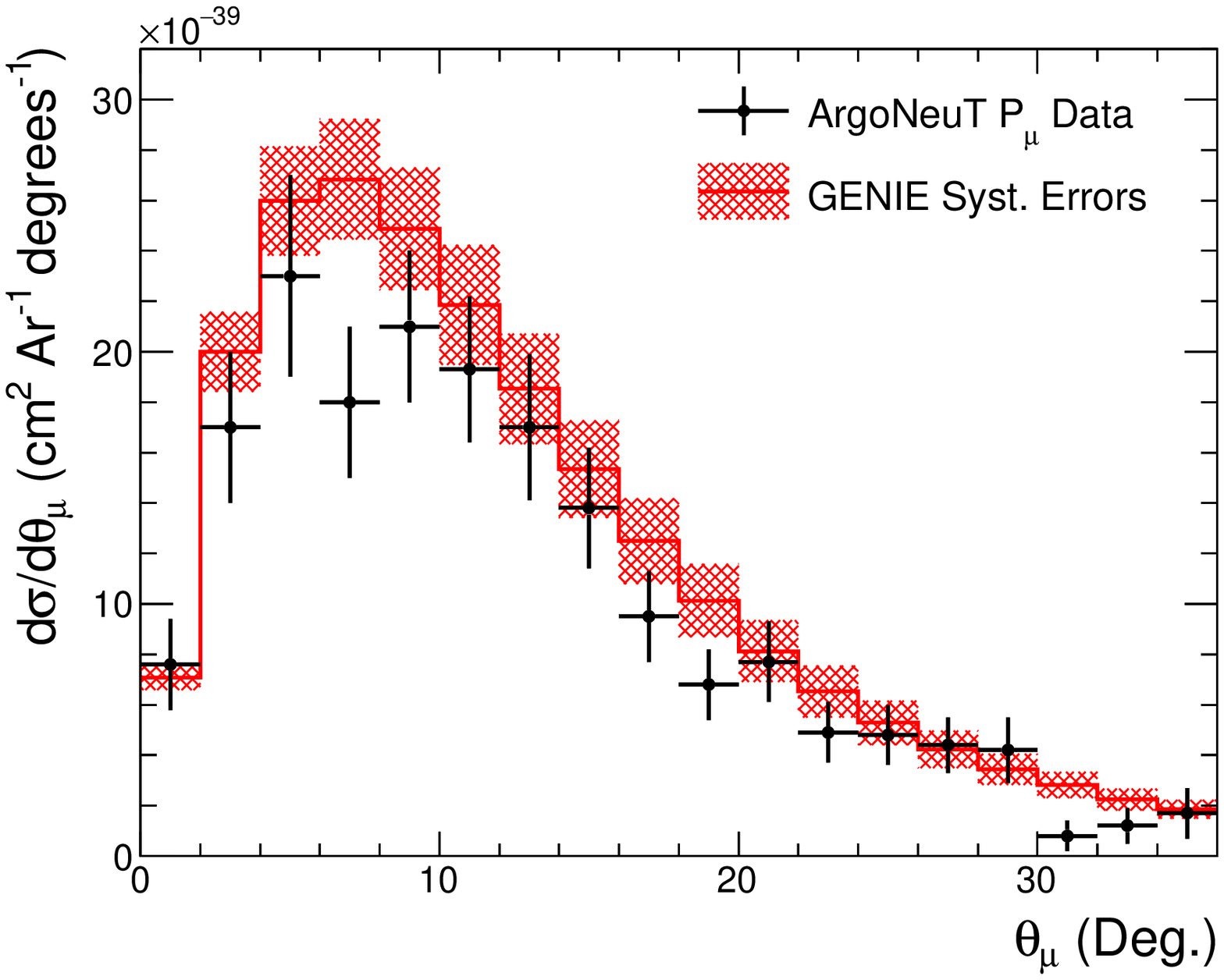}
    \caption{Muon angle w.r.t incoming beam direction ($\theta_{\mu}$)}
  \end{subfigure}
  \caption{Default GENIE v2.12.0 prediction for the ArgoNeut $\bar{\nu}_{\mu}$--$^{40}$Ar CC-inclusive distributions~\cite{Acciarri:2014isz}, with error bands produced with 500 throws of all GENIE reweighting parameters according to their nominal uncertainties.}
  \label{fig:gausbinthrows}
\end{figure}

A realistic use case for \emph{nuissyst} is shown in Figure~\ref{fig:gausbinthrows}, where 500 throws of the default GENIE v2.12.0 cross-section uncertainties have been used to build up a $1\sigma$ error band for the ArgoNeuT $\bar{\nu}$--$^{40}$Ar CC-inclusive dataset~\cite{Acciarri:2014isz}. This functionality enables the user to investigate whether a supplied generator cross-section uncertainty agrees with any particular dataset and aids in robust parameter error inflation studies. A histogram is saved in the output file which shows the level of data-MC agreement for all datasets included in the comparison. In such a case, a $\chi^{2}$ statistic is calculated as shown in Figure~\ref{fig:toymcthrows}. 

\subsection{Parameter fitting: \emph{nuismin}}
NUISANCE was originally designed to compare and tune the NEUT generator predictions to external datasets to provide cross-section uncertainties for T2K analyses. The \emph{nuismin} application fits any number of reweightable parameters to any combination of measurements, and uses ROOT minimisation libraries to minimise the test-statistic with respect to the parameters specified. ROOT provides multiple linear and non-linear scan methods which can be chosen at runtime. By default, the MIGRAD steepest gradient descent algorithm from the MINUIT package is used~\cite{James:1975dr}. It is also possible to modify the test-statistic for each dataset used in the minimisation by overloading the base class function. The fit parameters are specified in the card file, and it is possible to fix parameters, set fit boundaries and define starting values.

\begin{figure}
\centering
  \begin{subfigure}{0.49\columnwidth}
    \includegraphics[width=\textwidth]{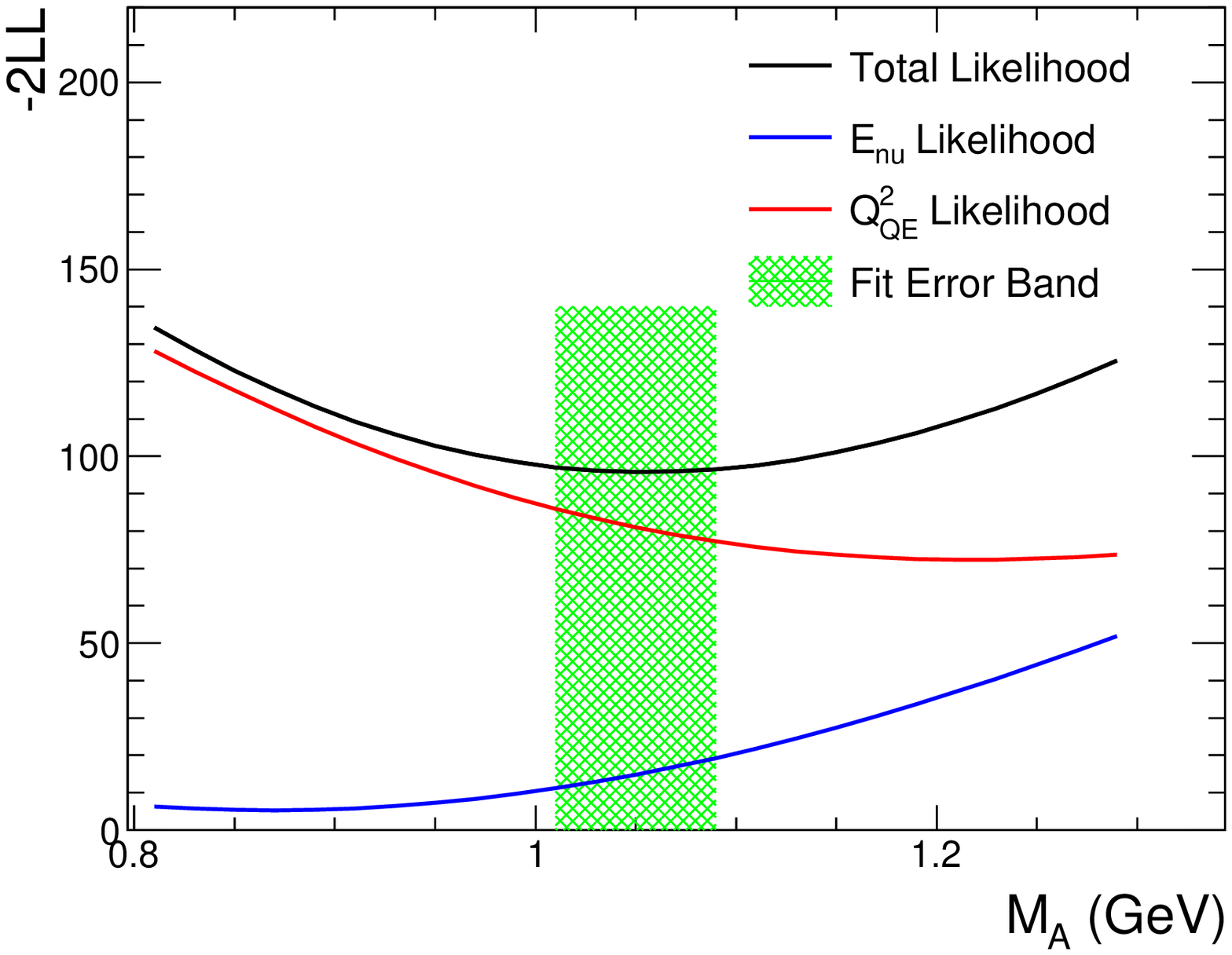}
    \caption{$\chi^{2}$ scan across the $M_{\mathrm{A}}$ parameter space}
  \end{subfigure}
  \begin{subfigure}{0.49\columnwidth}
    \includegraphics[width=\textwidth]{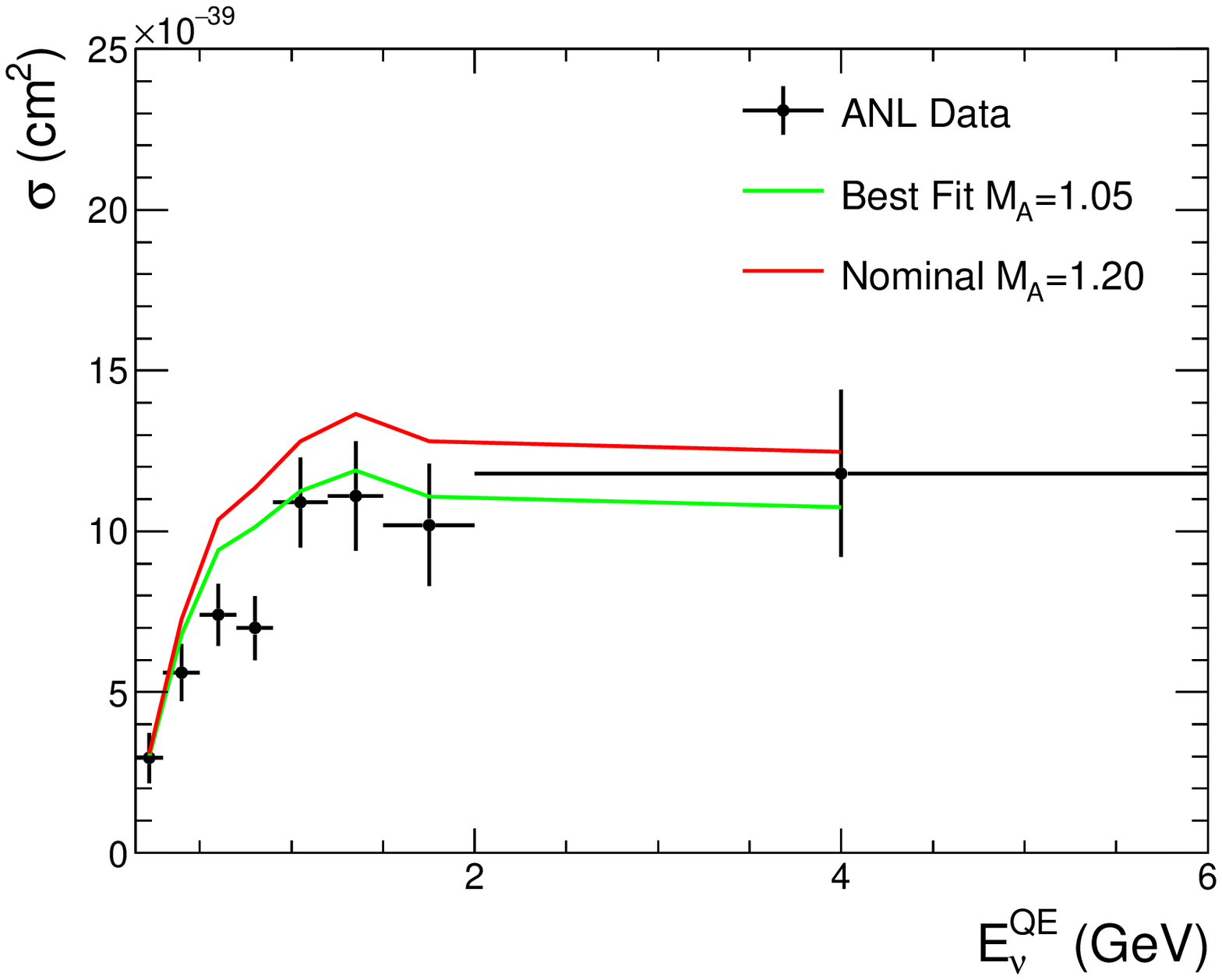}
    \caption{Nominal and best fit $E_{\nu}$ distribution}
  \end{subfigure}
\caption{Fit of the nominal NEUT v5.3.6 CCQE model to ANL CCQE $E_{\nu}$ and $N_{evt}(Q^2)$ distributions~\cite{Barish:1977qk} where the only free parameter in the axial mass $M_{\mathrm{A}}$. The $\chi^{2}$ contributions from each distribution to their joint likelihood is shown as a function of $M_{\mathrm{A}}$, and the best fit error is indicated. The best fit and nominal distributions are compared to the $E_{\nu}$ distribution.}
\label{fig:chi2scan}
\end{figure}

At each iteration of the fit, NUISANCE recalculates weights on an event-by-event basis using the relevant generator's reweighting libraries with the parameter variations requested by the minimisation algorithm. Weighted histograms are filled for all specified samples and the new test-statistic for the reweighted prediction is calculated and used to inform the minimisation algorithm. The output includes the nominal and best fit histograms, information about the best fit parameters and correlations between them, and basic information about the fit, such as the best fit $\chi^{2}$ and the number of iterations. Parameter error estimation is determined by the minimiser in ROOT. An example of a simple fit procedure is shown in Figure~\ref{fig:chi2scan}, where the NEUT CCQE model is fit simultaneously to ANL CCQE $\sigma(E_\nu)$ and $N_{evt}(Q^2)$ data~\cite{Barish:1977qk}. The joint likelihood is the sum of the likelihoods provided by each of the samples, which are treated independently by NUISANCE. The nominal and best fit distributions are also shown, and the error on the fitted parameter is indicated.

\begin{figure}
  \centering
  \centering
  \begin{subfigure}{0.49\columnwidth}
    \includegraphics[width=\textwidth]{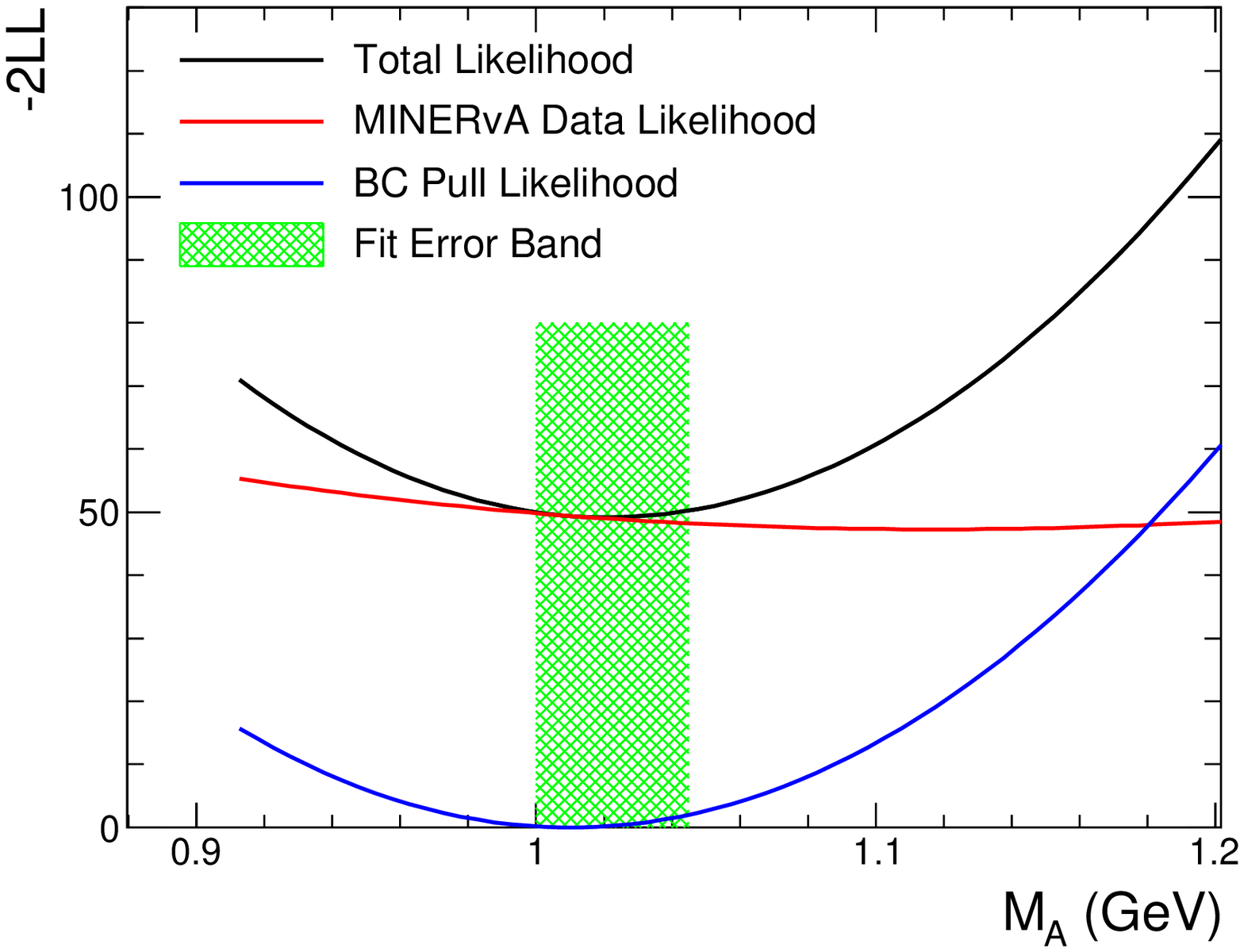}
    \caption{$\chi^{2}$ scan across the $M_{\mathrm{A}}$ parameter space}
  \end{subfigure}
  \begin{subfigure}{0.49\columnwidth}
    \includegraphics[width=\textwidth]{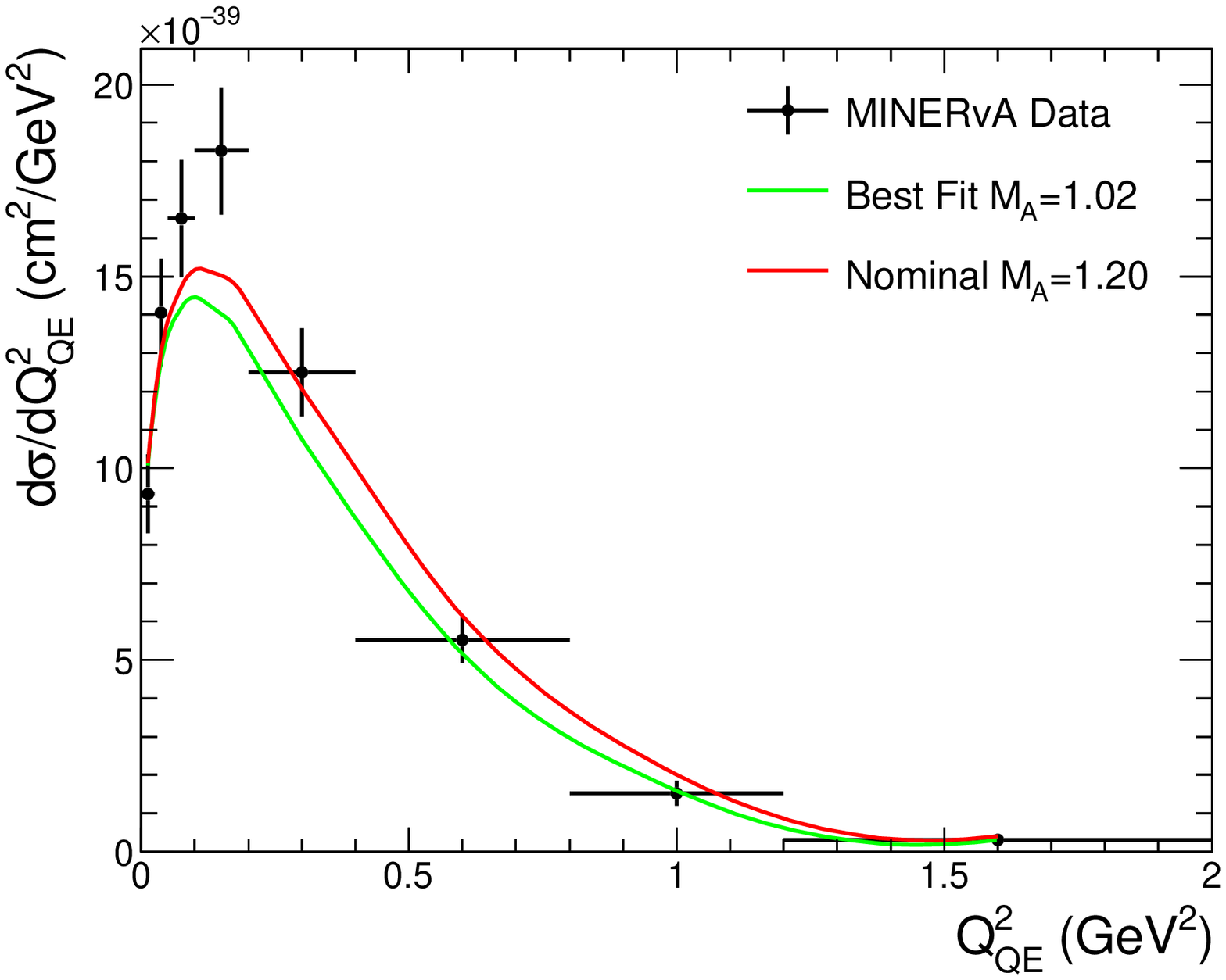}
    \caption{Nominal and best fit \minerva distribution}
  \end{subfigure}
  \caption{Fit of NEUT v5.3.6 to \minerva $\nu_{\mu}$--CH CCQE data~\cite{Fiorentini:2013ezn} with a prior constraint set by a fit to ANL CCQE data, shown in Figure~\ref{fig:chi2scan}.}
  \label{fig:chi2pulls}
  
\end{figure}

Penalty terms on parameters can also be introduced in \emph{nuismin}. The penalties can be applied with a correlation by supplying a covariance matrix. The output parameter covariance matrix of a previous \emph{nuismin} fit is also supported as a penalty term in a subsequent fit. For example, Figure~\ref{fig:chi2pulls} shows a fit to \minerva $\nu_{\mu}$--CH CCQE data~\cite{Fiorentini:2013ezn} where the only free parameter is the axial mass, $M_{\mathrm{A}}$. In this fit, the result of the free-nucleon fit to ANL data shown in Figure~\ref{fig:chi2scan} has been used as a prior constraint on $M_{\mathrm{A}}$ which contributes a penalty to the fit. The contribution from the $\chi^{2}$ from the \minerva data and the ANL prior is indicated, and it is clear that the \minerva data favours a higher $M_{\mathrm{A}}$, contesting the ANL prior. It is also possible to use the output of the \emph{nuismin} fit as an input to most other NUISANCE applications. For example, after running a fit to MiniBooNE CC1$\pi^+$ data, it might be desirable to produce error bands showing the effect of the uncertainty on T2K CC1$\pi^+$ data, for which the \emph{nuissyst} application can be used.

\FloatBarrier
\section{Summary}
NUISANCE is a flexible tool for comparing all commonly used neutrino event generators with published cross-section data. It provides a common ground for comparing generators as well as testing and tuning model parameters. It has already proven an invaluable tool for T2K studies of cross-section parameters~\cite{Wilkinson:2016wmz}, and has been made open access in the hope that it will prove useful to the wider community. A number of different possible usage cases have been identified for NUISANCE:
\begin{itemize}
  \item users who wish to make a comprehensive range of well-validated generator comparisons to new cross-section datasets without having to be familiar with all of the generators;
  
  \item users who wish to validate their cross-section parametrisation and error budget with a variety of historical cross-section data, or test new parameters against them;
  
  \item users who wish to tune and select default cross-section models for a given generator to a wide variety of cross-section data for cross-section and oscillation experiments;
  
  \item users who are interested in evaluating systematic uncertainties for systematics by comparing predictions of multiple generators.
\end{itemize}

This paper provides a number of examples to demonstrate the types of analyses which are straight-forward to perform with NUISANCE. Further documentation, usage examples and guidance can be found at \url{nuisance.hepforge.org}. We welcome code contributions, collaboration and new members.

\acknowledgments
The authors would like to thank the members of the T2K collaboration for the help and support when developing NUISANCE. We would also like to thank the authors of the NEUT, NuWro, GiBUU, and GENIE generators for being responsive to questions regarding the interface to their software. L. Pickering would like to extend special thanks to U. Mosel for his help validating the NUISANCE GiBUU integration. We thank the \minerva, \mb and T2K collaborations for assistance in understanding their results and helping us to validate the output from NUISANCE. We would like to thank HepForge for hosting the NUISANCE framework. We acknowledge the support of MEXT, Japan; 
National Science Centre (NCN), Poland; 
MINECO and ERDF funds, Spain; 
SNSF and SER, Switzerland; 
STFC, UK; 
and DOE, USA. KM would like to acknowledge the support of the DOE under award number 145568 and the Alfred P. Sloan Foundation.

\bibliography{nuisreference}
\bibliographystyle{unsrt}








\end{document}